\documentclass{aa}
\usepackage{graphicx}
\usepackage{psfig,lscape}
\def\kms{$\rm km\;s^{-1}$}
\def\dg{^\circ}

\def\kmsmpc{$\rm km\;s^{-1}\;Mpc^{-1}$}

\def\vs{$v_\star$}
\def\Vs{$V_\star$}
\def\vg{$v_{\rm g}$}
\def\Vg{$V_{\rm g}$}
\def\ss{$\sigma_\star$}
\def\ssc{$\sigma_\star(0)$}
\def\sg{$\sigma_{\rm g}$}

\def\ha{H$\alpha$}
\def\nii{[N~{\small II}]}

\def\niipg{[N~{\small II}]$\,\lambda6583$}

\def\siipg{[S~{\small II}]$\,\lambda\lambda6716, 6731$}

\begin{document}

\title{Ionized gas and stellar kinematics of seventeen nearby spiral galaxies
\thanks{Based on observations carried out at the European Southern
  Observatory, La Silla (Chile) (ESO 56.A-0684 and
  57.A-0569).}$^{\bf,}$
\thanks{Tables 3 and 4 are only available in electronic
 form at the CDS via anonymous ftp to cdsarc.u-strasbg.fr (130.79.128.5)
 or via http://cdsweb.u-strasbg.fr/Abstract.html.}}

\author{A. Pizzella\inst{1}
 \and E.M. Corsini\inst{1}
 \and J.C. Vega-Beltr\'an\inst{2}
 \and F. Bertola\inst{1} }

\offprints{pizzella@pd.astro.it}

\institute{
Dipartimento di Astronomia, Universit\`a di Padova, vicolo
dell'Osservatorio~2, I-35122 Padova, Italy \and Instituto Astrof\'\i
sico de Canarias, Calle V\'\i a L\'actea s/n, E-38200 La Laguna, Spain}

\date{\today}

\titlerunning{Kinematics of late-type spirals}
\authorrunning{Pizzella et al.}

\abstract{Ionized gas and stellar kinematics have been measured along
the major axes of seventeen nearby spiral galaxies of intermediate to late
morphological type. We discuss the properties of each sample galaxy
distinguishing between those characterized by regular or peculiar
kinematics. In most of the observed galaxies ionized gas rotates
more rapidly than stars and have a lower velocity dispersion, as  is to be
expected if the gas is confined in the disc and supported by rotation
while the stars are mostly supported by dynamical pressure. In a few
objects, gas and stars show almost the same rotational velocity and low
velocity dispersion, suggesting that their motion is dominated by
rotation.
Incorporating the spiral galaxies studied by Bertola et al. (1996),
Corsini et al.\ (1999, 2003) and Vega Beltr\'an et al.\ (2001) we have
compiled a sample of 50 S0/a--Scd galaxies, for which the major-axis
kinematics of the ionized gas and stars have been obtained with the same
spatial ($\approx1''$) and spectral  ($\approx 50$ \kms) resolution,
and measured with the same analysis techniques. This allowed us to
address the frequency of counterrotation in spiral galaxies. It
turns out that less than $12\%$ and less than $8\%$ (at the $95\%$
confidence level) of the sample galaxies host a counterrotating
gaseous and stellar disc, respectively. The comparison with S0
galaxies suggests that the retrograde acquisition of small amounts of
external gas gives rise to counterrotating gaseous discs only in
gas-poor S0s, while in gas-rich spirals the newly acquired gas is
swept away by the pre-existing gas. Counterrotating gaseous and
stellar discs in spirals are formed only from the retrograde
acquisition of large amounts of gas exceeding that of pre-existing
gas, and subsequent star formation, respectively.
}

\maketitle

\section{Introduction}
\label{sec:introduction}

Studying the interplay between ionized gas and stellar kinematics
allows us to address different topics concerning the dynamical structure
of disc galaxies and to constrain the processes leading to their
formation and evolution.
These topics include the study of the mass distribution of luminous and dark
matter (see Sofue \& Rubin 2001 for a review), the ubiquity of
supermassive black holes and their relationship with the large-scale
properties of the host galaxies (see Merritt \& Ferrarese 2001 for a
review), the discovery of kinematically decoupled components (see
Bertola \& Corsini 1999 for a review), the origin of disc heating
(Merrifield, Gerssen \& Kuijken 2001 and references therein), and the
presence of pressure-supported ionized gas in bulges (Bertola et al.\
1995; Cinzano et al.\ 1999).

All these issues will greatly benefit from a survey devoted to the
comparative measurements of ionized gas and stellar kinematics in S0s
and spiral galaxies. Since these  are available only for a
limited number of objects, in the past years we began a scientific
programme aimed at deriving detailed velocity curves and velocity
dispersion radial profiles of ionized gas and stars along the major
axes of disc galaxies (Bertola et al.\ 1995, 1996; Corsini et al.\ 1999,
2003; Vega Beltr\'an et al.\ 2001; Funes et al.\ 2002) to be used for
mass modelling (Corsini et al.\ 1999; Cinzano et al.\ 1999; Pignatelli et
al.\ 2001).

This paper is organized as follows. An overview of the properties of
the sample galaxies as well as the spectroscopic observations and data
analysis are presented in Section \ref{sec:observations}.
The resulting ionized gas and stellar kinematics are given and
interpreted in Section \ref{sec:kinematics}. In particular, we derived
the central velocity dispersion of stars of all the sample galaxies
with the aim of studying the relationship between that the disc
circular velocity and bulge velocity dispersion (Ferrarese 2002;
Baes et al.\ 2003) in a forthcoming paper (but see also
Pizzella et al.\ 2003).
The fraction of spiral galaxies hosting a counterrotating component is
estimated in Section \ref{sec:counterrotation} by analysing the
major-axis kinematics of all the spiral galaxies we have observed in recent
years.
Our conclusions are discussed in Section \ref{sec:conclusions}.

\begin{table*}[t]
\caption[]{Parameters of the sample galaxies}
\begin{center}
\begin{footnotesize}
\begin{tabular}{lllrrcrrrrrcc}
\hline
\noalign{\smallskip}
\multicolumn{1}{c}{Object} &
\multicolumn{2}{c}{Type} &
\multicolumn{1}{c}{$B_T$} &
\multicolumn{1}{c}{PA} &
\multicolumn{1}{c}{$i$} &
\multicolumn{1}{c}{$V_{\odot}$} &
\multicolumn{1}{c}{$D$} &
\multicolumn{1}{c}{Scale} &
\multicolumn{1}{c}{$R_{25}$} &
\multicolumn{1}{c}{$M_{B_T}^0$} &
\multicolumn{2}{c}{$R_{\rm last}/R_{25}$} \\

\multicolumn{1}{c}{} &
\multicolumn{1}{c}{} &
\multicolumn{1}{c}{} &
\multicolumn{1}{c}{} &
\multicolumn{1}{c}{} &
\multicolumn{1}{c}{} &
\multicolumn{1}{c}{} &
\multicolumn{1}{c}{} &
\multicolumn{1}{c}{} &
\multicolumn{1}{c}{} &
\multicolumn{1}{c}{} &
\multicolumn{1}{c}{[stars]} &
\multicolumn{1}{c}{[gas]} \\

\multicolumn{1}{c}{[name]} &
\multicolumn{1}{c}{[RSA]} &
\multicolumn{1}{c}{[RC3]} &
\multicolumn{1}{c}{[mag]} &
\multicolumn{1}{c}{[\degr]} &
\multicolumn{1}{c}{[\degr]} &
\multicolumn{1}{c}{[\kms]} &
\multicolumn{1}{c}{[Mpc]} &
\multicolumn{1}{c}{[pc$/''$]} &
\multicolumn{1}{c}{[$''$]} &
\multicolumn{1}{c}{[mag]} &
\multicolumn{1}{c}{} &
\multicolumn{1}{c}{} \\

\multicolumn{1}{c}{$^{(1)}$} &
\multicolumn{1}{c}{$^{(2)}$} &
\multicolumn{1}{c}{$^{(3)}$} &
\multicolumn{1}{c}{$^{(4)}$} &
\multicolumn{1}{c}{$^{(5)}$} &
\multicolumn{1}{c}{$^{(6)}$} &
\multicolumn{1}{c}{$^{(7)}$} &
\multicolumn{1}{c}{$^{(8)}$} &
\multicolumn{1}{c}{$^{(9)}$} &
\multicolumn{1}{c}{$^{(10)}$} &
\multicolumn{1}{c}{$^{(11)}$} &
\multicolumn{1}{c}{$^{(12)}$} &
\multicolumn{1}{c}{$^{(13)}$} \\
\noalign{\smallskip}
\hline
\noalign{\smallskip}
\object{NGC 210}  & Sb(rs)  & SABb(s)     & 11.60 & 165 & 49 & 1650 & 23.2 & 113 & 150 & $-20.5$ & 0.3&0.8 \\
\object{NGC 615}  & Sb(r)   & SAb(rs)     & 12.47 & 155 & 67 & 1849 & 25.7 & 125 & 108 & $-20.3$ & 0.7&0.9 \\
\object{NGC 1620} & ---     & SABbc(rs)   & 13.08 &  25 & 70 & 3509 & 46.1 & 224 &  86 & $-21.1$ & 0.7&1.2 \\
\object{NGC 2590} & ---     & SAbc(s):    & 13.94 &  77 & 72 & 4960 & 63.4 & 308 &  67 & $-21.0$ & 0.5&1.0 \\
\object{NGC 2708} & ---     & SABb(s)pec? & 12.80 &  25 & 60 & 1984 & 23.5 & 114 &  78 & $-19.4$ & 0.6&0.8 \\
\object{NGC 2815} & Sb(s)   & SBb(r):     & 12.81 &  10 & 72 & 2535 & 30.0 & 145 & 104 & $-21.0$ & 0.4&1.0 \\
\object{NGC 3054} & SBbc(s) & SABb(r)     & 12.35 & 118 & 52 & 2430 & 28.5 & 138 & 114 & $-20.3$ & 0.5&0.9 \\
\object{NGC 3200} & Sb(r)   & SABc(rs):   & 12.83 & 169 & 73 & 3526 & 43.4 & 211 & 125 & $-21.5$ & 0.5&0.9 \\
\object{NGC 3717} & Sb(s)   & SAb:sp      & 12.24 &  33 & 81 & 1748 & 19.6 &  95 & 180 & $-20.5$ & 0.4&0.5 \\
\object{NGC 4682} & Sc(s)   & SABcd(s)    & 13.14 &  83 & 61 & 2344 & 28.8 & 140 &  77 & $-20.0$ & 0.3&0.8 \\
\object{NGC 5530} & Sc(s)   & SAbc(rs)    & 11.79 & 127 & 62 & 1092 & 11.7 &  57 & 125 & $-19.4$ & 0.6&0.9 \\
\object{NGC 6118} & Sc(s)   & SAcd(s)     & 12.42 &  58 & 65 & 1580 & 21.4 & 104 & 140 & $-20.4$ & 0.6&0.8 \\
\object{NGC 6878} & Sc(r)   & SAb(s)      & 13.45 & 125 & 40 & 5821 & 77.3 & 375 &  48 & $-21.3$ & 0.6&1.0 \\
\object{NGC 6925} & Sbc(r)  & SAbc(s)     & 12.07 &   5 & 75 & 2783 & 37.7 & 183 & 134 & $-21.9$ & 0.6&0.9 \\
\object{NGC 7083} & Sb(s)   & SAbc(s)     & 11.87 &   5 & 53 & 3093 & 39.7 & 192 & 116 & $-21.5$ & 0.5&1.0 \\
\object{NGC 7412} & Sc(rs)  & SBc(s)      & 11.88 &  65 & 42 & 1712 & 22.7 & 110 & 116 & $-20.1$ & 0.4&0.8 \\
\object{NGC 7531} & Sbc(r)  & SABbc(r)    & 12.04 &  15 & 67 & 1591 & 20.9 & 102 & 134 & $-20.2$ & 0.4&0.6 \\
\noalign{\smallskip}
\hline
\noalign{\smallskip}
\noalign{\smallskip}
\noalign{\smallskip}
\end{tabular}
\begin{minipage}{18cm}
NOTES -- $^{(2)}$Morphological classification from Sandage \& Tammann
         (1981, RSA hereafter).
$^{(3)}$Morphological classification from RC3.
$^{(4)}$Total observed blue magnitude from RC3 except for
         NGC~3954, NGC 4682 and NGC 5530 (LEDA).
$^{(5)}$Major-axis position angle from RC3.
$^{(6)}$Inclination, derived as
         $\cos^{2}{i}\,=\,(q^2-q_0^2)/(1-q_0^2)$. The observed axial ratio $q$
         is taken from RC3 and the intrinsic flattening $q_0=0.11$ has been
         assumed following Guthrie (1992).
$^{(7)}$Heliocentric velocity of the galaxy derived at the centre of
         symmetry of the rotation curve of the gas. $\Delta V_\odot = 10$
         \kms .
$^{(8)}$Distance obtained as $V_0/H_0$ with $H_0=75$ \kms\
         Mpc$^{-1}$ and $V_0$, the systemic velocity derived from $V_\odot$
         corrected for the motion of the Sun with respect to the Local Group,
         as in  RSA.
$^{(10)}$Radius of the 25 $B$ mag arcsec$^{-2}$ isophote derived as
          $R_{25} = D_{25}/2$ with $D_{25}$ from RC3.
$^{(11)}$Absolute total blue magnitude corrected for
          inclination and extinction from RC3.
$^{(12)}$Radial extension of the stellar rotation curve
          in units of $R_{25}$ from this paper.
$^{(13)}$Radial extension of the ionized gas rotation curve
          in units of $R_{25}$ from this paper.
\end{minipage}
\end{footnotesize}
\end{center}
\label{tab:sample_properties}
\end{table*}

\section{Sample selection, spectroscopic observations and data reduction}
\label{sec:observations}

\subsection{Sample selection}
\label{sec:sample}

\begin{figure}
   \centering \includegraphics[width=8.5cm]{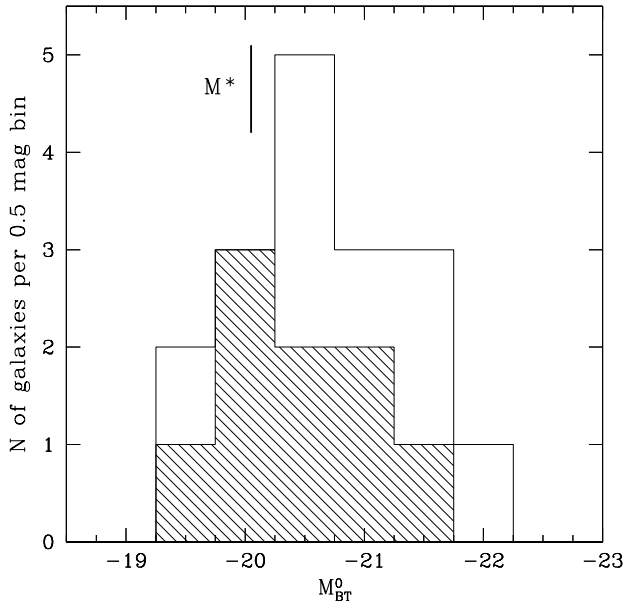}
\caption{Absolute magnitude distribution for the sample galaxies.
  A line marks $M_{B_T}^0 = -20.05$, which corresponds to M$^\ast$ for
  spiral galaxies as derived by Marzke et al. (1998) and assuming $H_0
  = 75$ \kmsmpc . The dashed region identifies galaxies classified
  barred or weakly barred in RC3.}
\label{fig:histogram}
\end{figure}

All the observed galaxies are bright ($B_T\leq13.5$) and nearby
($V_\odot < 5900$ \kms) with an intermediate-to-high inclination
($40\dg \leq i \leq 85\dg$). The Hubble morphological type of the
sample galaxies ranges from Sb to Scd (de Vaucouleurs et al. 1991, RC3
hereafter) with eight unbarred, seven weakly barred  and two strongly barred
galaxies (RC3). The galaxies have been chosen for their strong emission
lines. An overview of the basic properties of the sample galaxies is given
in Table~\ref{tab:sample_properties}. The distribution of their
absolute magnitudes, which brackets the M$^\ast$ value for spiral
galaxies (Marzke et al. 1998, for $H_0 = 75$ \kmsmpc) is shown in
Figure \ref{fig:histogram}.

\subsection{Spectroscopic observations}
\label{sec:spectroscopy}

The spectroscopic observations of our sample galaxies were carried out
at the ESO in La Silla with the 1.52 m ESO telescope on 1996 January 22--27 
 (run 1 hereafter), and on 1996 August 9--14  (run 2 hereafter).

The telescope mounted the Boller \& Chievens Spectrograph. The grism
No. 26 with 1200 $\rm grooves\,mm^{-1}$ blazed at 5730 \AA\ was used
in the first order in combination with the $2\farcs5\times4\farcm2$
slit and the Loral CCD No. 24 with $2048\,\times\,2048$ pixels of
$15\,\times\,15$ $\rm \mu m^2$. The wavelength range between 4940 \AA\
and 6940 \AA\ was covered with a reciprocal dispersion of 0.98 \AA\
pixel$^{-1}$, which guarantees adequate oversampling of the
instrumental broadening function. Indeed, the instrumental resolution,
obtained by measuring the width of emission lines of a comparison
spectrum after the wavelength calibration, was $2.59$ \AA\ (FWHM).
This corresponds to an instrumental velocity dispersion of $\sigma =
1.10$ \AA\ (i.e.\ $\approx70$ and $\approx50$ \kms\ at the blue and
red edges of the spectra, respectively). The angular sampling was
$0\farcs81$ pixel$^{-1}$.

At the beginning of each exposure, the slit was centred on the galaxy
nucleus using the guiding TV camera and aligned along the galaxy major
axis. Details of the slit position and spectra exposure times are
given in Table \ref{tab:log}. Comparison lamp exposures were obtained
before and/or after each object integration to allow an accurate
wavelength calibration.  Quartz lamp and twilight sky flat-fields were
used to map pixel-to-pixel sensitivity variations and large scale
illumination patterns.  A dozen of late G and early K stars were
observed with the same set up to serve as templates in measuring the
stellar kinematics.  The value of the seeing FWHM during the different
observing runs ranged between $1\farcs0$ and $2\farcs0$, as measured by
fitting a two-dimensional Gaussian to the guide star.

\begin{table}[ht!]
\caption{Log of spectroscopic observations}
\begin{tabular}{lccr}
\hline
\noalign{\smallskip}
\multicolumn{1}{c}{Object} &
\multicolumn{1}{c}{Run}&
\multicolumn{1}{c}{t$_{\rm exp}$} &
\multicolumn{1}{c}{P.A.}\\
\noalign{\smallskip}
\multicolumn{1}{c}{[name]} &
\multicolumn{1}{c}{} &
\multicolumn{1}{c}{[s]} &
\multicolumn{1}{c}{[$\dg$]} \\
\noalign{\smallskip}
\hline
\noalign{\smallskip}
NGC~210  &  2  & $5\times3600$ & 160 \\
NGC~615  &  2  & $7\times3600$ & 155 \\
NGC~1620 &  1  & $7\times3600$ &  25 \\
NGC~2590 &  1  & $4\times3600$ &  77 \\
NGC~2708 &  1  & $5\times3600$ &  20 \\
NGC~2815 &  1  & $5\times3600$ &  10 \\
NGC~3054 &  1  & $5\times3600$ & 118 \\
NGC~3200 &  1  & $4\times3600$ & 169 \\
NGC~3717 &  1  & $3\times3600$ &  33 \\
NGC~4682 &  1  & $3\times3600$ &  85 \\
NGC~5530 &  2  & $3\times3600$ & 127 \\
NGC~6118 &  2  & $5\times3600$ &  58 \\
NGC~6878 &  2  & $6\times3600$ & 125 \\
NGC~6925 &  2  & $8\times3600$ &   5 \\
NGC~7083 &  2  & $6\times3600$ &   5 \\
NGC~7412 &  2  & $6\times3600$ &  65 \\
NGC~7531 &  2  & $5\times3600$ &  15 \\
\noalign{\smallskip}
\hline
\end{tabular}
\label{tab:log}
\end{table}

\subsection{Routine data reduction}
\label{sec:datareduction}

The spectra were bias subtracted, flat-field corrected, cleaned for
cosmic rays and wavelength calibrated using standard
MIDAS\footnote{MIDAS is developed and maintained by the European
Southern Observatory.} routines.  Cosmic rays were identified by
comparing the counts in each pixel with the local mean and standard
deviation (as obtained from the Poisson statistics of the photons
knowing the gain and readout noise of the detector), and then
corrected by interpolating a suitable value. In the central regions
(typically for $|r|\la5''$) cosmic rays have been removed by manually
editing the spectra.

The instrumental resolution as a function of wavelength was derived as
the Gaussian FWHMs measured for several unblended arc lamp lines
distributed over the whole spectral range of a wavelength-calibrated
comparison spectrum. Finally, the spectra of the same galaxy were
aligned and co-added using the centers of their stellar continua as
reference. In the resulting spectra, the contribution from the sky was
determined by interpolating along the outermost $10''$--$20''$ at the
edges of the slit, where galaxy light was negligible. The sky level
was then subtracted.  Since the spatial extension of the emission
lines can exceed that of the stellar continuum, different ranges for
sky subtraction have been considered in ionized gas and stellar
kinematic measurements.  In few cases the galaxy emission lines
extended through the whole slit length.  The sky level was derived by
scaling the sky frame derived for a different galaxy to match the
intensity of the relevant night sky emission lines.

\subsection{Measuring ionized gas and stellar kinematics}
\label{sec:measuring}

The ionized gas kinematics was measured by the simultaneous Gaussian
fit of the emission lines present in the spectra (namely \niipg, \ha,
and \siipg). The galaxy continuum was removed from the spectra, as was done
for measuring the stellar kinematics. We fitted 
 a Gaussian to each emission line in each row of the
continuum-subtracted spectrum,
assuming them to have the same line-of-sight velocity (\vg), and
velocity dispersion (\sg). Velocities and velocity dispersions were
corrected for heliocentric velocity and instrumental FWHM,
respectively. An additional absorption Gaussian was added to the
fit to take into account for the presence of the \ha\ absorption line
and the flux ratio of the \nii\ lines have been fixed at 1:3. Far from
the galaxy centre (for $|r|\ga10''$) we averaged adjacent spectral
rows in order to increase the signal-to-noise ratio of the relevant
emission lines.
We checked for a few galaxies that the error in the kinematic parameter
determination derived by means of Monte Carlo simulations did not
differ significantly from the formal errors given as output by the
least-squares fitting routine.  We therefore decided to assume the
latter as error bars on the gas kinematics.
For each galaxy we derive the heliocentric systemic velocity,
$V_\odot$, as the velocity of the centre of symmetry of the
ionized gas rotation curve (Table~\ref{tab:sample_properties}).

The stellar kinematics were measured from the galaxy absorption
features present in the wavelength range running from 5050 \AA\ to
5550 \AA\ and centred on the Mg line triplet
($\lambda\lambda\,5164,5173,5184$
\AA).
We used the Fourier correlation quotient method (FCQ, Bender 1990)
following the prescriptions of Bender, Saglia \& Gerhard (1994).  The
spectra were binned along the spatial direction to obtain a nearly
constant signal-to-noise ratio larger than 20 per resolution element.
The galaxy continuum was removed row by row by fitting a fourth to
sixth order polynomial as in Bender et al.  (1994). The K1\ III star
\object{HR 5777} was adopted as a kinematic template. This allowed us
to derive, for each spectrum, the line-of-sight stellar velocity
(\vs), and velocity dispersion (\ss) by fitting a Gaussian to the
line-of-sight velocity distribution (LOSVD) at each radius. Velocities
were corrected for heliocentric velocity.
We derived errors for the stellar kinematics from photon statistics and
CCD read-out noise, calibrating them by Monte Carlo simulations as
done by Gerhard et al. (1998). These errors do not take into account
possible systematic effects due to any template mismatch.

The ionized gas and stellar kinematics of all the sample galaxies are
tabulated in Tables 3 and 4, respectively, and plotted in Figure
\ref{fig:kinematics}.

\section{The ionized gas and stellar kinematics}
\label{sec:kinematics}

\subsection{Velocity curves and velocity dispersion
profiles of ionized gas and stars}

\begin{figure*}[t!]
\begin{minipage}[t]{8.5cm}
\vspace{0pt}
\psfig{figure=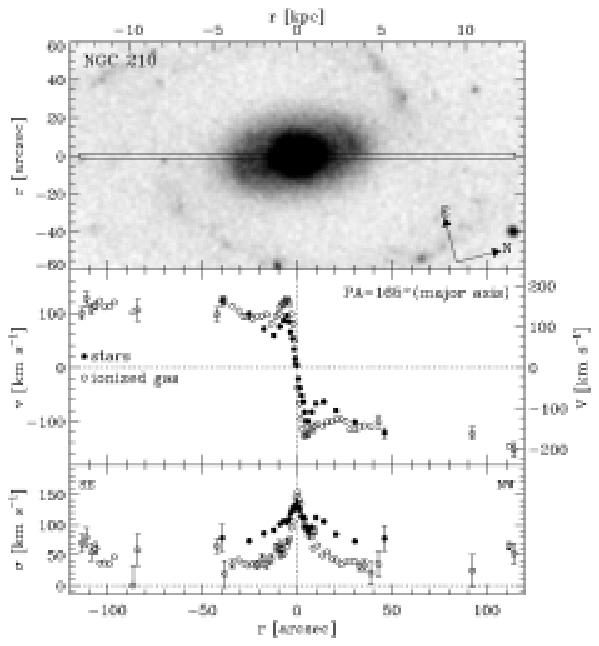,width=8.5cm}
\end{minipage}
\hspace*{0.5cm}
\begin{minipage}[t]{8.5cm}
\vspace{0pt}
\psfig{figure=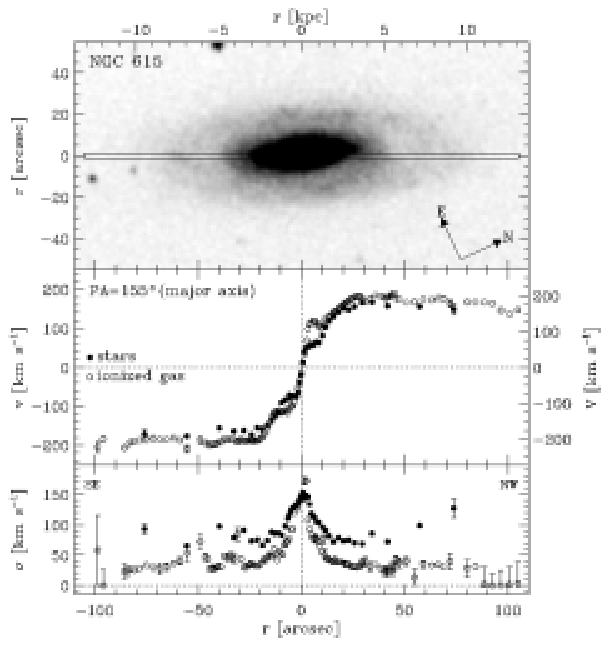,width=8.5cm}
\end{minipage}
\vspace*{0.cm}
\begin{minipage}[t]{8.5cm}
\vspace{20pt}
\psfig{figure=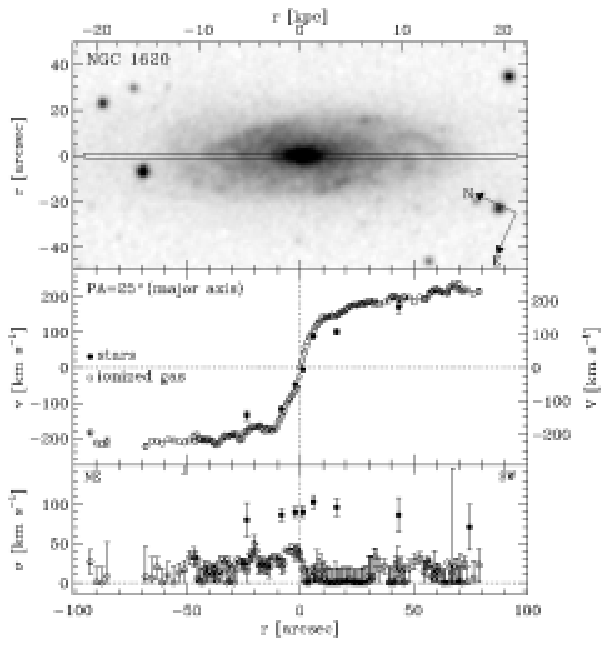,width=8.5cm}
\end{minipage}
\hspace*{0.5cm}
\begin{minipage}[t]{8.5cm}
\vspace{20pt}
\psfig{figure=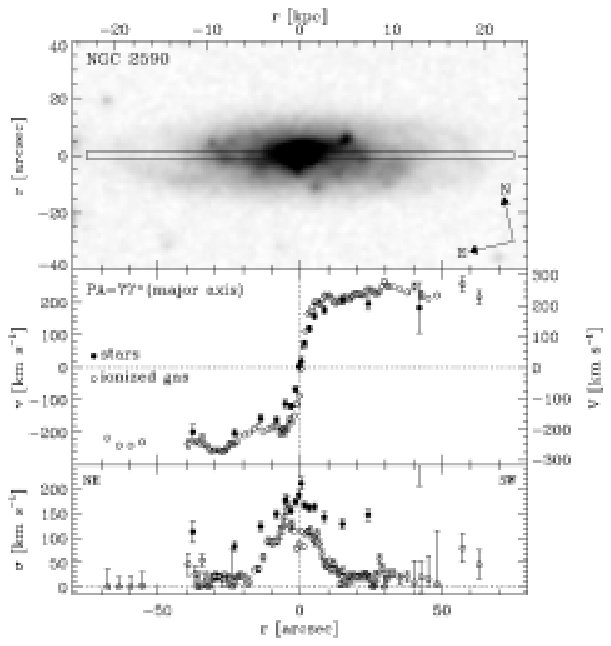,width=8.5cm}
\end{minipage}
\caption{Ionized gas ({\it open circles\/}) and stellar
  ({\it filled circles\/}) kinematics measured along the
  major axes of the sample galaxies. Error bars smaller than their symbols are not
  plotted.}
\label{fig:kinematics}
\end{figure*}

The resulting ionized gas and stellar kinematics of all our sample
galaxies are shown in Figure \ref{fig:kinematics}.  For each object the
plot is organized as it follows:

\begin{enumerate}
\item In the upper panel we display the galaxy image obtained
  from the Digitized Sky Survey. The galaxy image has been rotated and
  the slit position has been plotted to allow a better comparison
  between morphological and kinematic properties. The slit width and
  length correspond to those adopted in obtaining the spectra.

\item In the middle panel we plot the velocity curves of the gaseous
  (open circles) and stellar (filled circles) components. The velocity
  scale on the left  of the plot indicates the observed velocity
  after subtraction of the systemic velocity given in
  Table~\ref{tab:sample_properties}.  The velocity scale on the right
   indicates the rotation velocity after the deprojection for the
  galaxy inclination given in Table~\ref{tab:sample_properties}. Error
  bars are not plotted when smaller than symbols. The position angle
  of the slit is specified.

\item In the lower panel we plot in radial profiles the velocity
  dispersion of the gaseous (open circles) and stellar (filled
  circles) components. Error bars are not plotted when smaller
  than their symbols.

\end{enumerate}

The kinematic measurements obtained for the sample galaxies extend typically
 to  $\approx0.9\;R_{25}$ for the gaseous and
$\approx0.5\;R_{25}$ for the stellar component. The curves are
generally regular and symmetric about the centre.

\addtocounter{figure}{-1}
\begin{figure*}[t!]
\begin{minipage}[t]{8.5cm}
\vspace{0pt}
\psfig{figure=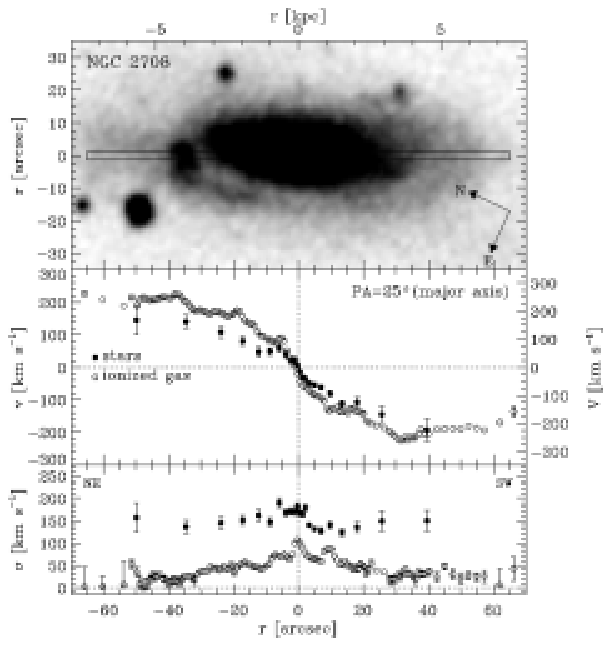,width=8.5cm}
\end{minipage}
\hspace*{0.5cm}
\begin{minipage}[t]{8.5cm}
\vspace{0pt}
\psfig{figure=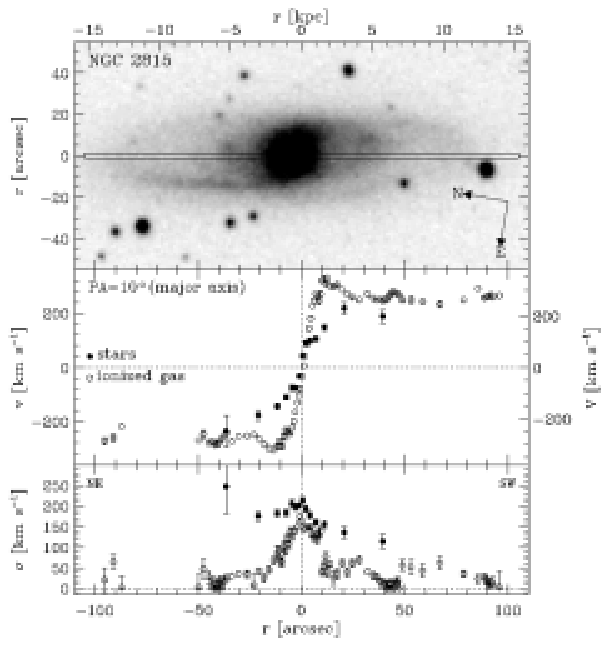,width=8.5cm}
\end{minipage}
\vspace*{0.cm}
\begin{minipage}[t]{8.5cm}
\vspace{20pt}
\psfig{figure=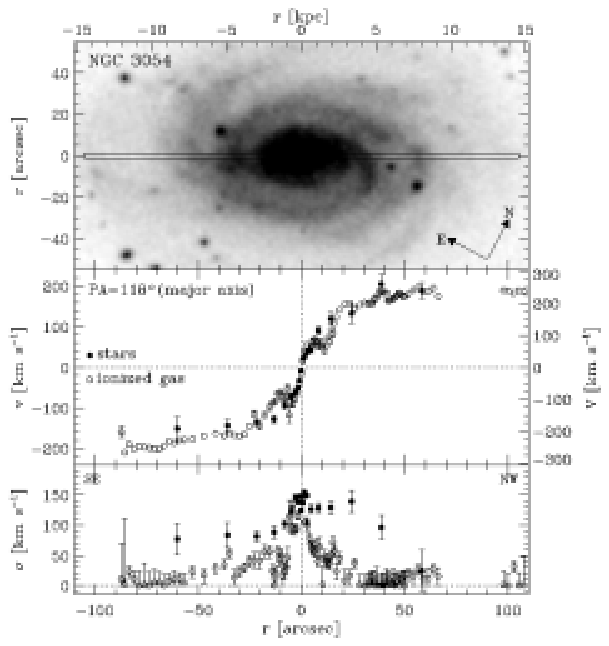,width=8.5cm}
\end{minipage}
\hspace*{0.5cm}
\begin{minipage}[t]{8.5cm}
\vspace{20pt}
\psfig{figure=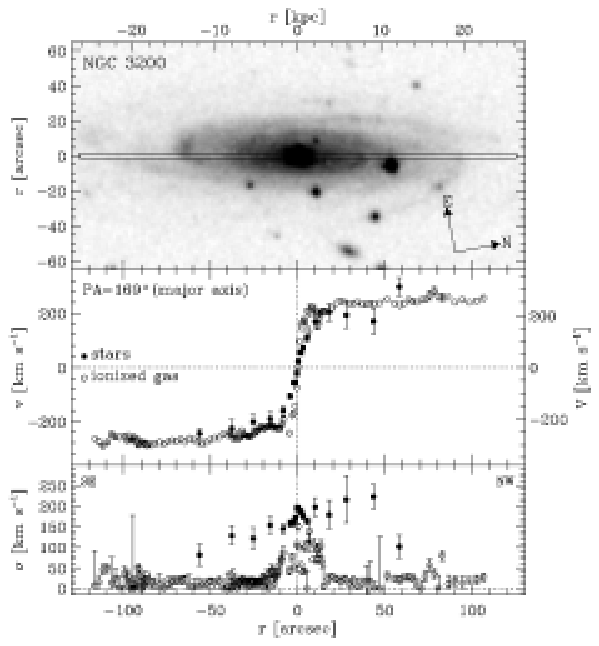,width=8.5cm}
\end{minipage}
\caption{(continue).}
\end{figure*}

\addtocounter{figure}{-1}
\begin{figure*}[t!]
\begin{minipage}[t]{8.5cm}
\vspace{0pt}
\psfig{figure=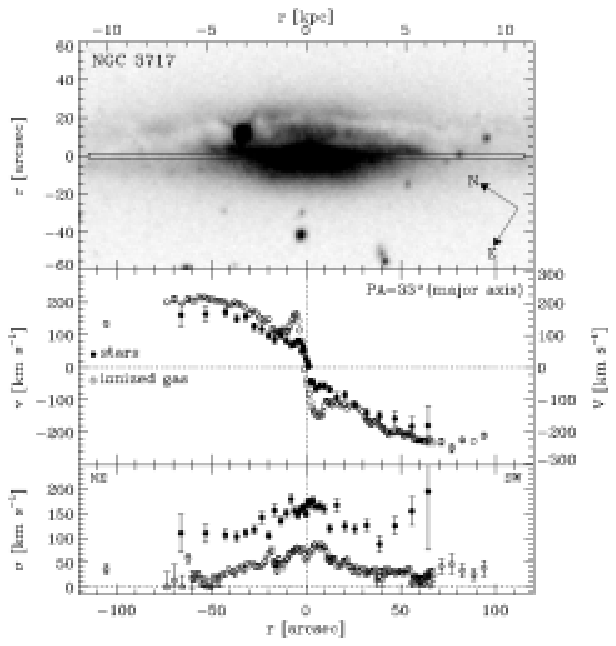,width=8.5cm}
\end{minipage}
\hspace*{0.5cm}
\begin{minipage}[t]{8.5cm}
\vspace{0pt}
\psfig{figure=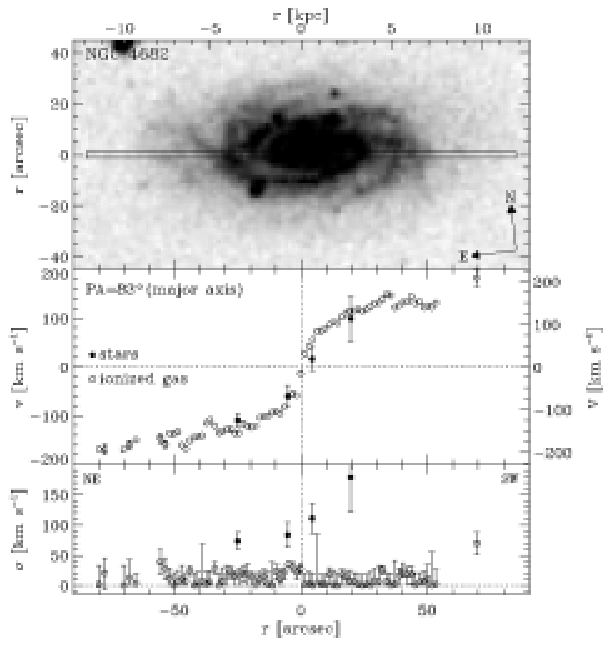,width=8.5cm}
\end{minipage}
\vspace*{0.cm}
\begin{minipage}[t]{8.5cm}
\vspace{20pt}
\psfig{figure=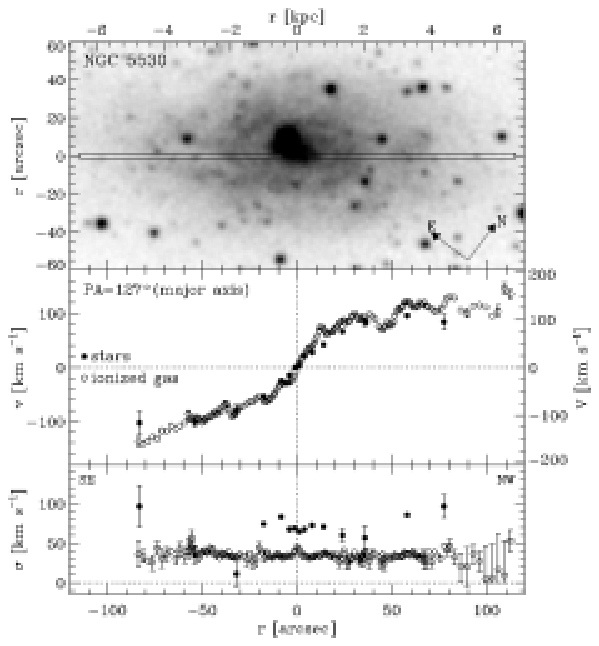,width=8.5cm}
\end{minipage}
\hspace*{0.5cm}
\begin{minipage}[t]{8.5cm}
\vspace{20pt}
\psfig{figure=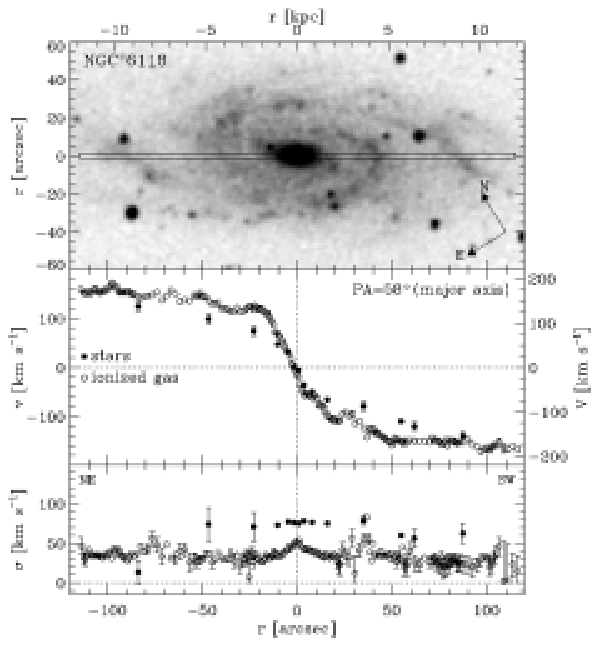,width=8.5cm}
\end{minipage}
\caption{(continue).}
\end{figure*}

\addtocounter{figure}{-1}
\begin{figure*}[t!]
\begin{minipage}[t]{8.5cm}
\vspace{0pt}
\psfig{figure=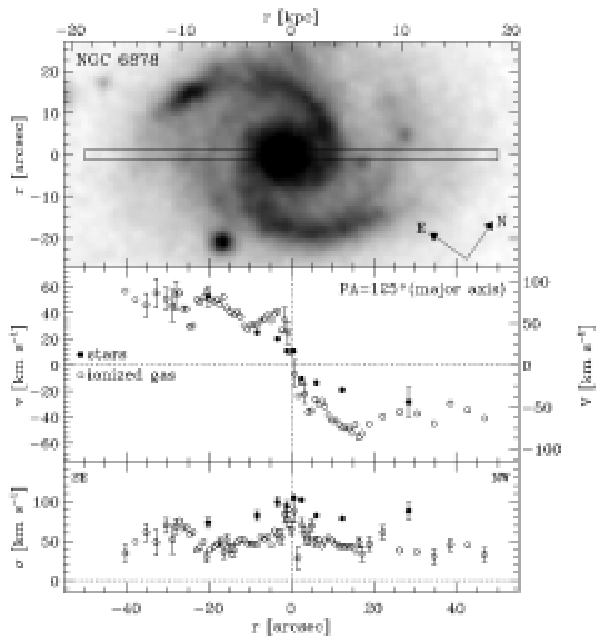,width=8.5cm}
\end{minipage}
\hspace*{0.5cm}
\begin{minipage}[t]{8.5cm}
\vspace{0pt}
\psfig{figure=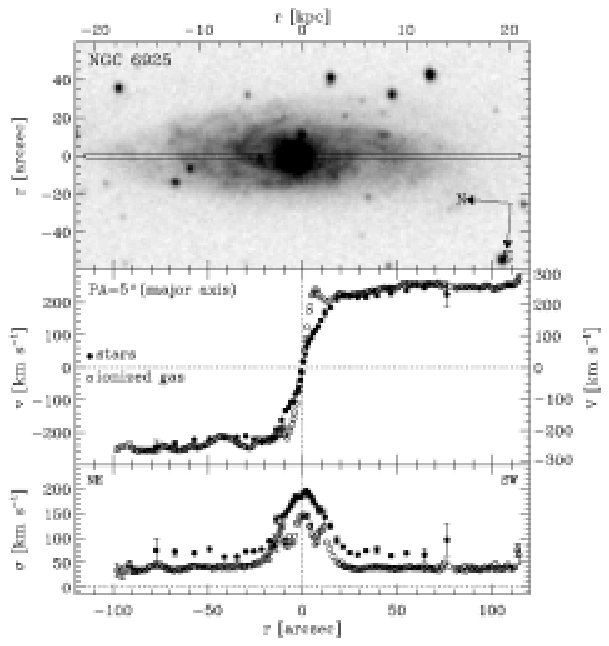,width=8.5cm}
\end{minipage}
\vspace*{0.cm}
\begin{minipage}[t]{8.5cm}
\vspace{20pt}
\psfig{figure=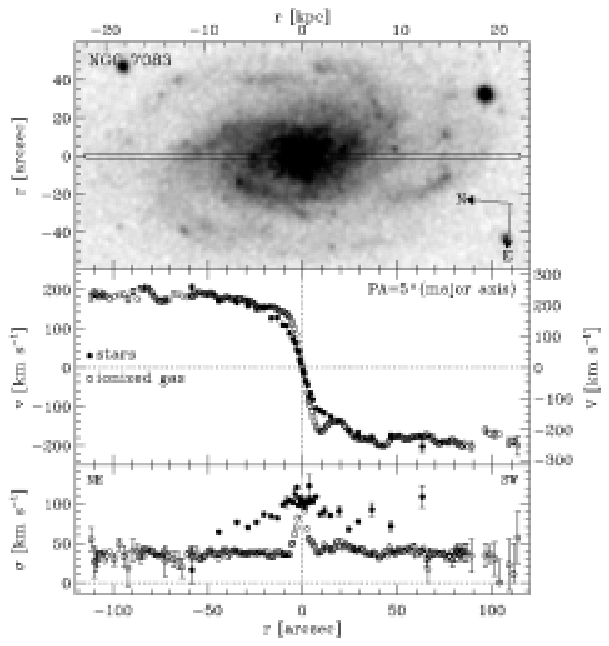,width=8.5cm}
\end{minipage}
\hspace*{0.5cm}
\begin{minipage}[t]{8.5cm}
\vspace{20pt}
\psfig{figure=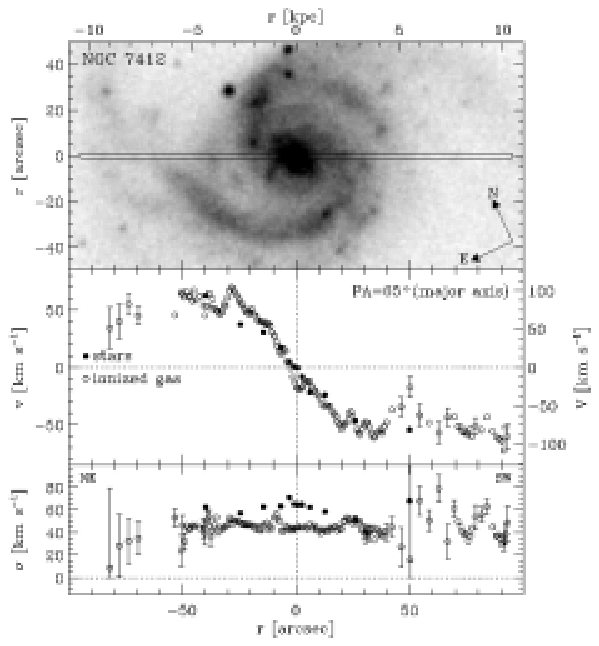,width=8.5cm}
\end{minipage}
\caption{(continue).}
\end{figure*}

\addtocounter{figure}{-1}
\begin{figure}[h!]
\vspace*{0.cm}
\begin{minipage}[t]{8.5cm}
\vspace{20pt}
\psfig{figure=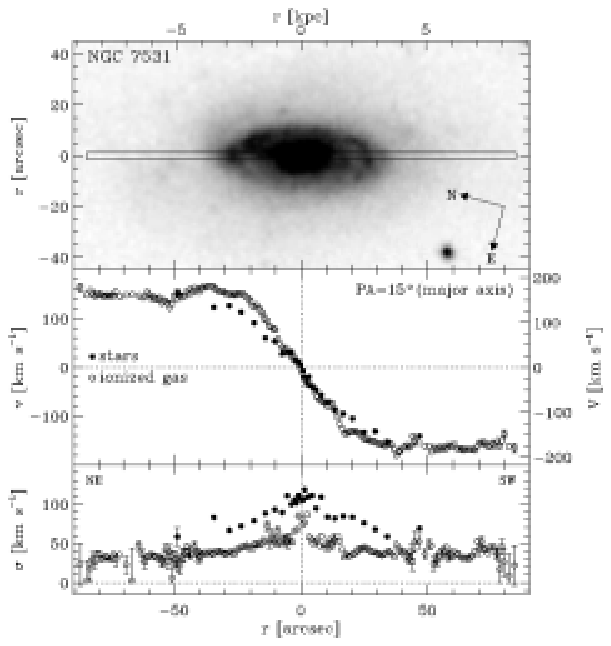,width=8.5cm}
\end{minipage}
\caption{(continue).}
\end{figure}

\subsection{Comparison with the literature}

The ionized gas and/or stellar kinematics of some of the sample
galaxies have been measured by other authors. We used these data sets
to assess the accuracy and reliability of our measurements.  In most
cases differences between different authors are due to slit centring
and/or positioning, different analysis techniques, or both (see Fisher
1997 for a discussion).

\begin{figure*}
\begin{minipage}[t]{8.5cm}
\vspace{0pt}
\includegraphics[width=8.5cm]{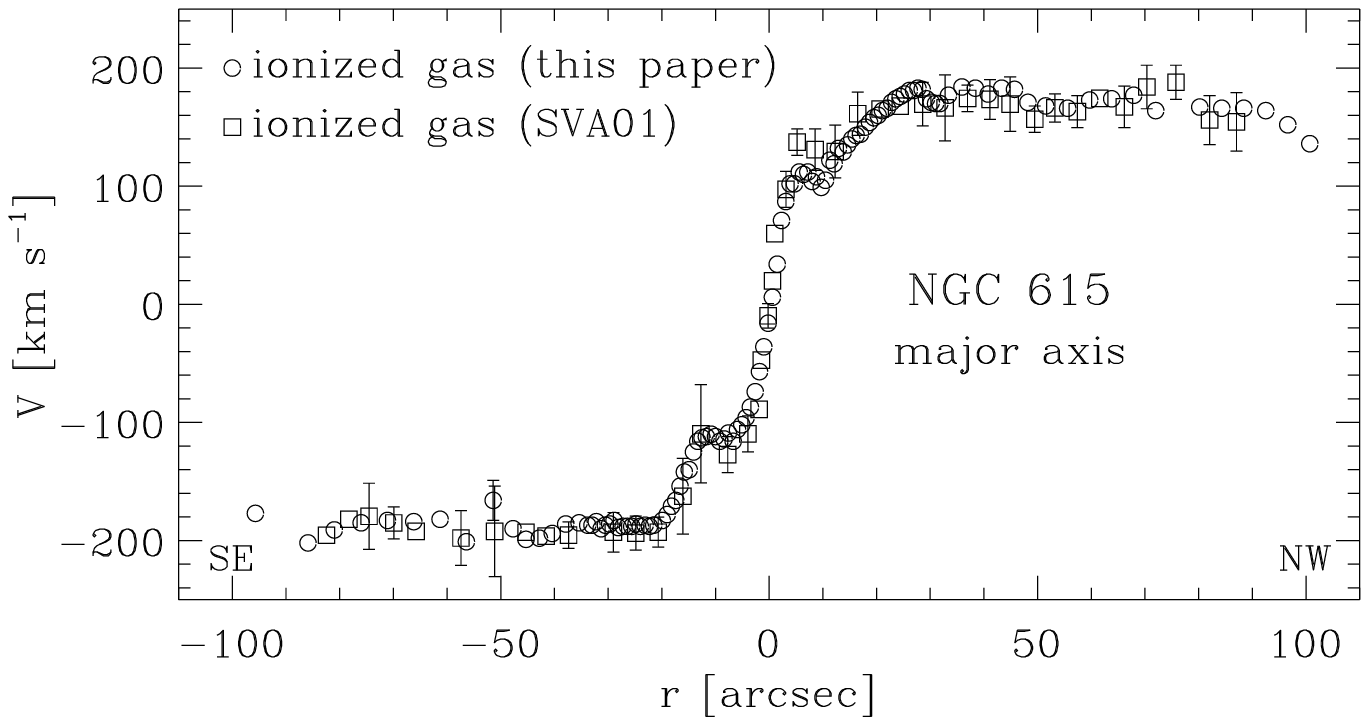}
\end{minipage}
\hspace*{0.5cm}
\begin{minipage}[t]{8.5cm}
\vspace{0pt}
\includegraphics[width=8.5cm]{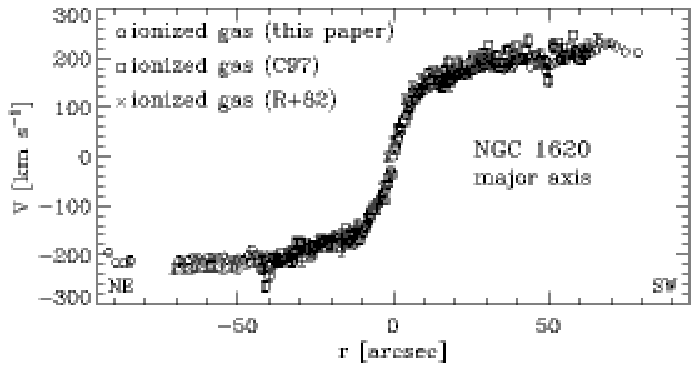}
\end{minipage}
\begin{minipage}[t]{8.5cm}
\vspace{0pt}
\includegraphics[width=8.5cm]{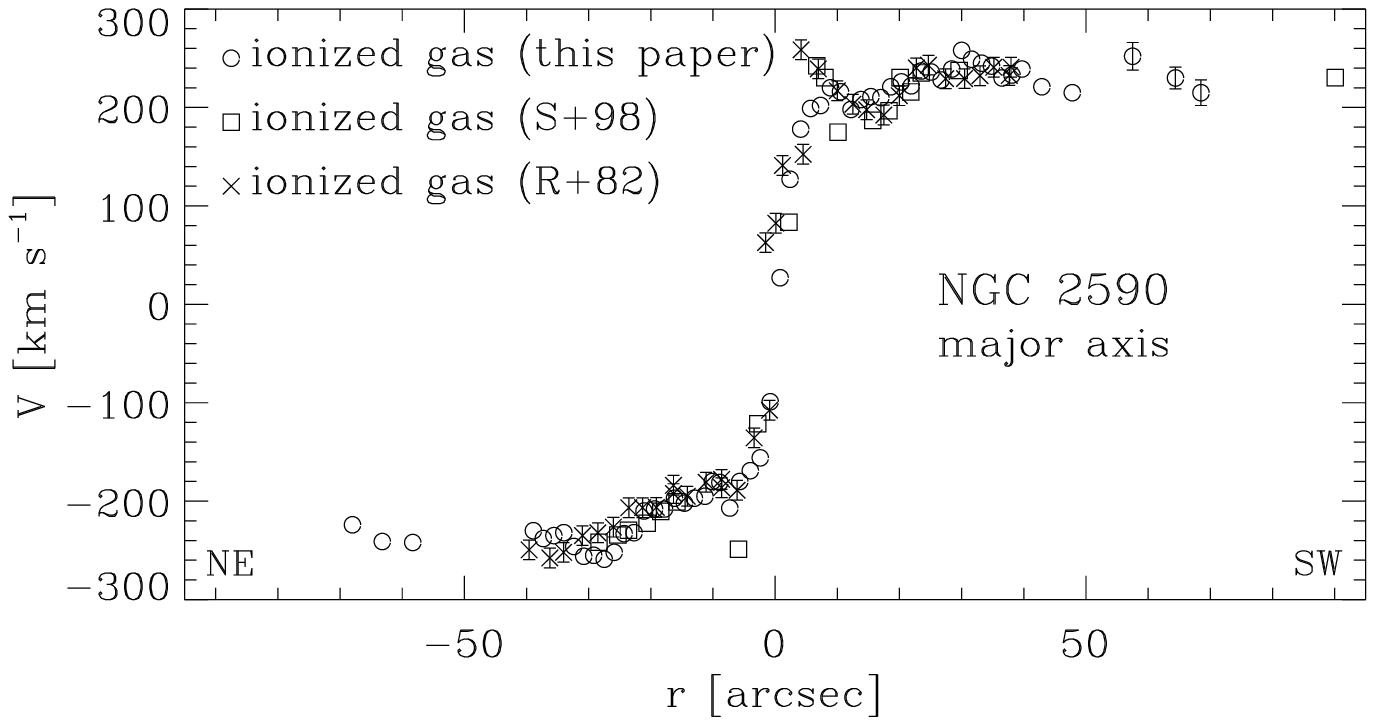}
\end{minipage}
\hspace*{0.5cm}
\begin{minipage}[t]{8.5cm}
\vspace{0pt}
\includegraphics[width=8.5cm]{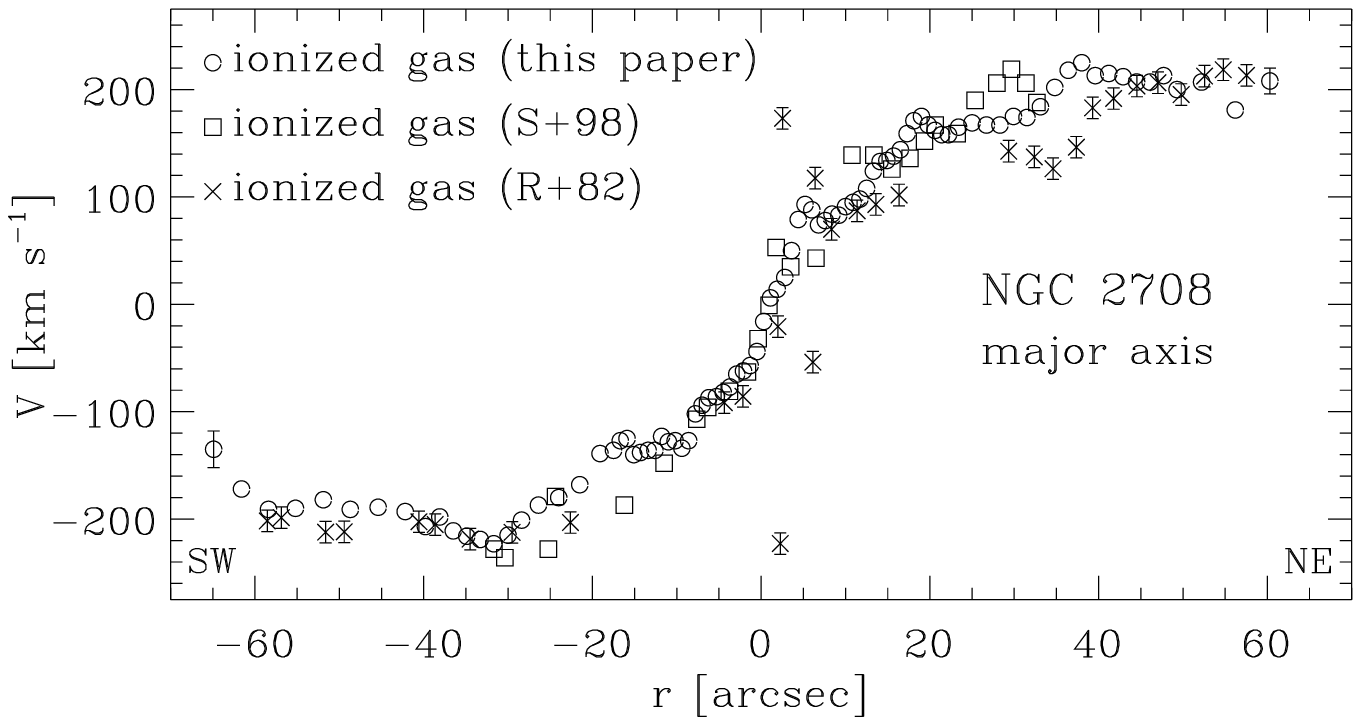}
\end{minipage}
\begin{minipage}[t]{8.5cm}
\vspace{0pt}
\includegraphics[width=8.5cm]{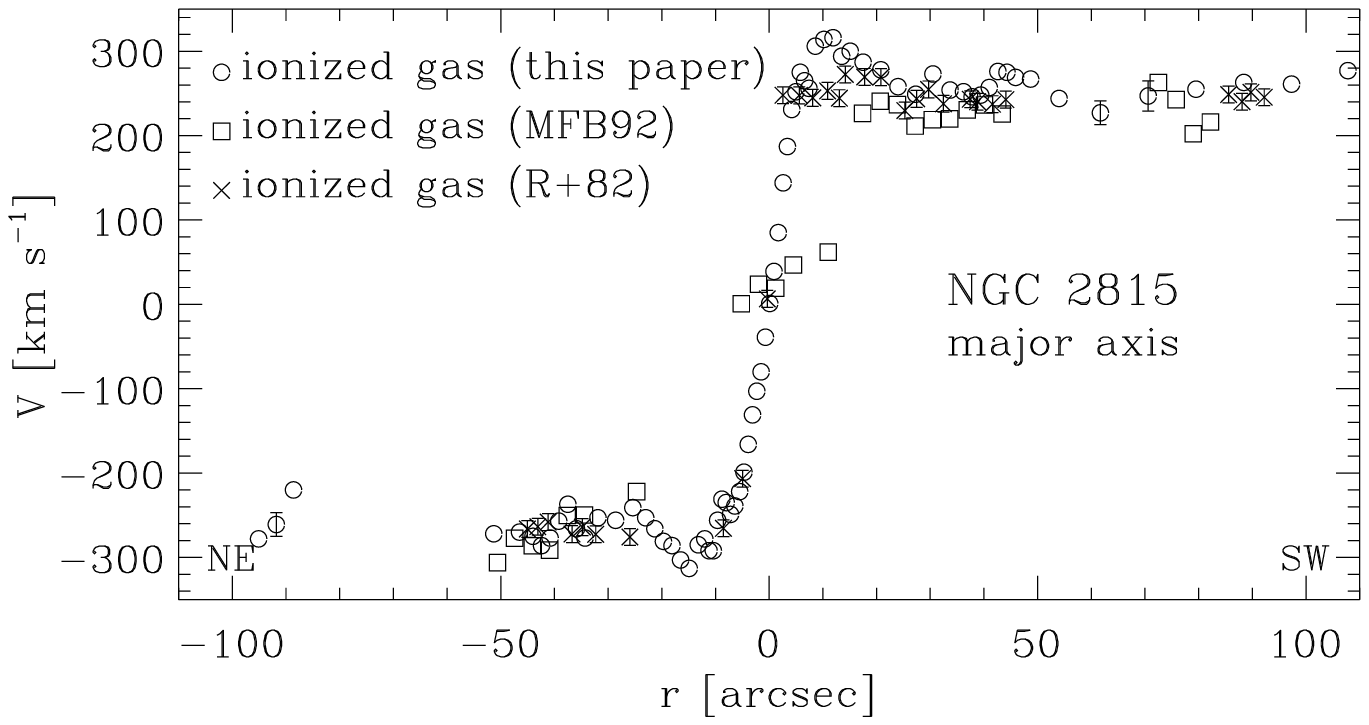}
\end{minipage}
\hspace*{0.5cm}
\begin{minipage}[t]{8.5cm}
\vspace{0pt}
\includegraphics[width=8.5cm]{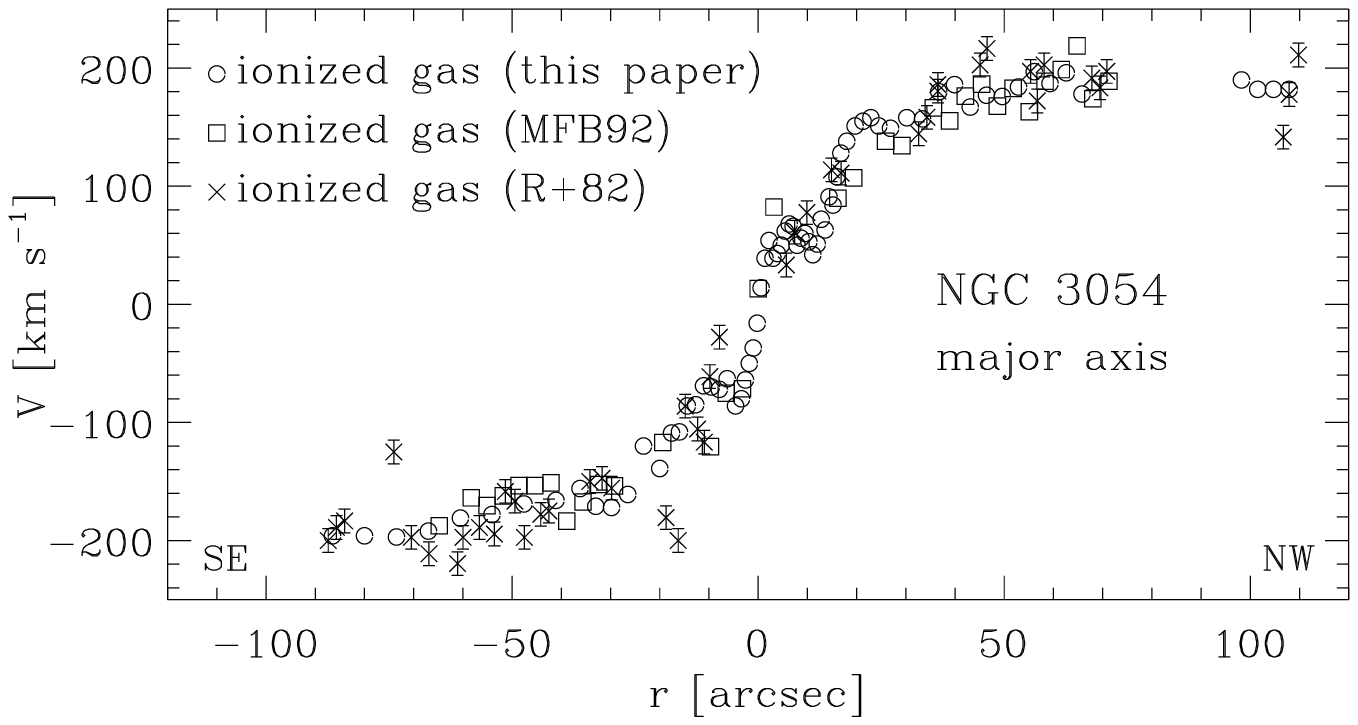}
\end{minipage}
\caption{The ionized gas velocities derived in this study for NGC
  615, NGC 1620, NGC 2708, NGC 2815 and NGC 3054 compared with those
  obtained by other authors: MFB92 = Mathewson, Ford \& Buchhorn
  (1992); R$+$82 = Rubin et al.\  (1982); SVA01 = Sil'chenko, Vlasyuk
  \& Alvarado (2001); C97 = Courteau (1997); S$+$98 = Sofue et
  al.\ (1998).}
\label{fig:gascomparison}
\end{figure*}

Ionized gas velocity curves have been already measured along the major
axes of NGC 615, NGC 1620, NGC 2708, NGC 2815, NGC 3054, NGC 3200, NGC
4682, NGC 6118, NGC 7083 and NGC 7531. However, these measurements
had previously been obtained with a lower spatial and/or velocity resolution
with respect to ours.  In Figure \ref{fig:gascomparison} we plot our
kinematic measurements compared to the available data.  The
agreement is good and the discrepancies ($\Delta V\la20$ \kms) are due
to the different spatial sampling and accuracy of velocity
measurements. Only in the case of NGC~7083 do we notice a systematic
difference ($\Delta V\approx80$ \kms) on the north-east side of the
rotation curve between our data and those of Rubin et al.\ (1982). We
attribute such a discrepancy to the wavelength calibration of Rubin et
al. (1982) since it is not present when we consider the data of
Mathewson, Ford  \& Buchhorn (1992).

\addtocounter{figure}{-1}
\begin{figure*}
\begin{minipage}[t]{8.5cm}
\vspace{0pt}
\includegraphics[width=8.5cm]{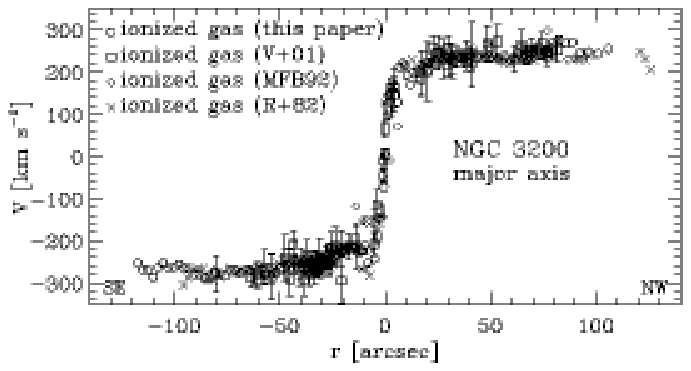}
\end{minipage}
\hspace*{0.5cm}
\begin{minipage}[t]{8.5cm}
\vspace{0pt}
\includegraphics[width=8.5cm]{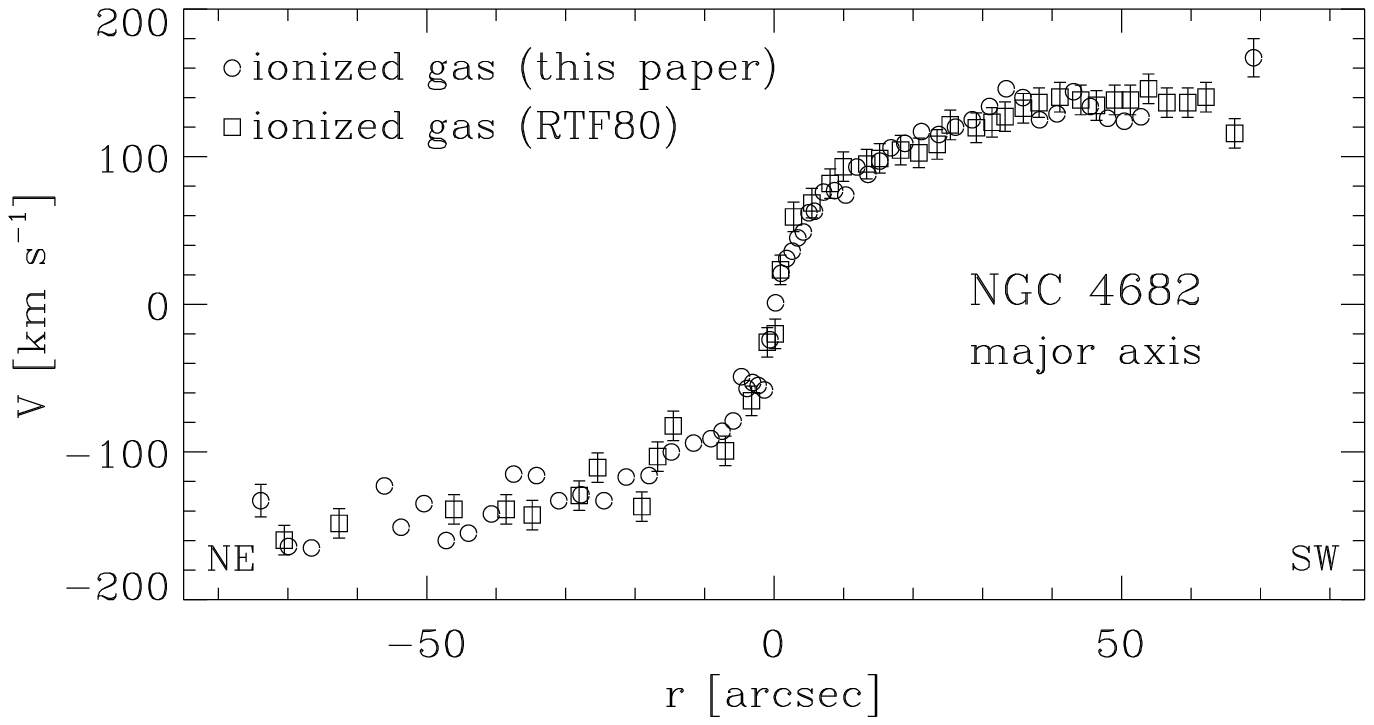}
\end{minipage}
\begin{minipage}[t]{8.5cm}
\vspace{0pt}
\includegraphics[width=8.5cm]{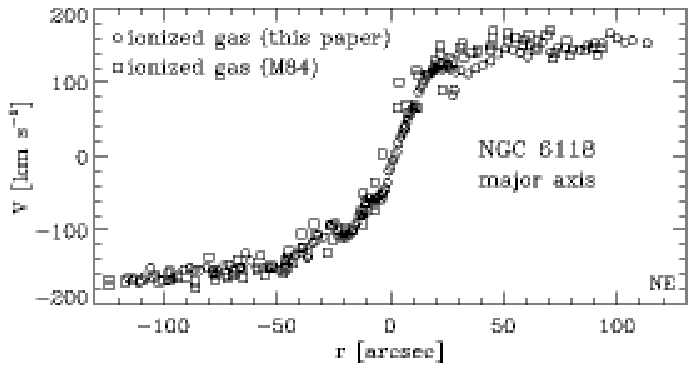}
\end{minipage}
\hspace*{0.5cm}
\begin{minipage}[t]{8.5cm}
\vspace{0pt}
\includegraphics[width=8.5cm]{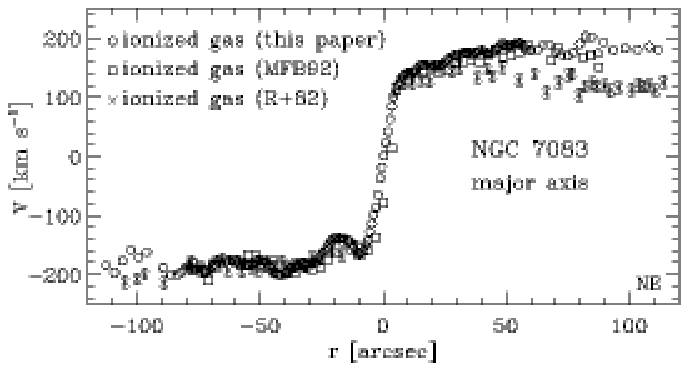}
\end{minipage}
\begin{minipage}[t]{8.5cm}
\vspace{0pt}
\includegraphics[width=8.5cm]{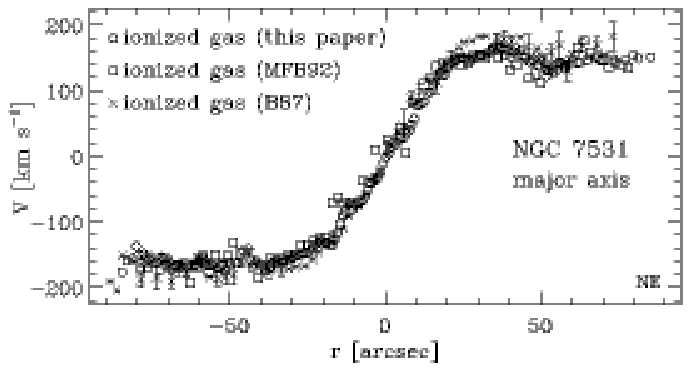}
\end{minipage}
\hspace*{0.5cm}
\caption{(continue). The ionized gas velocities derived in this study for
  NGC 3200, NGC 4682, NGC 6118, NGC 7083 and NGC 7531 compared with
  those obtained by other authors: V$+$01 = Vega Beltr{\'a}n et
  al.\ (2001).; MFB92 = Mathewson, Ford \& Buchhorn (1992); R$+$82 =
  Rubin et al.\ (1982); RTF80 = Rubin, Thonnard \& Ford (1980); B87 = Buta
  (1987); M84 = Meyssonnier (1984).}
\end{figure*}

\begin{figure*}
\begin{minipage}[t]{8.5cm}
\vspace{0pt}
\includegraphics[width=8.5cm]{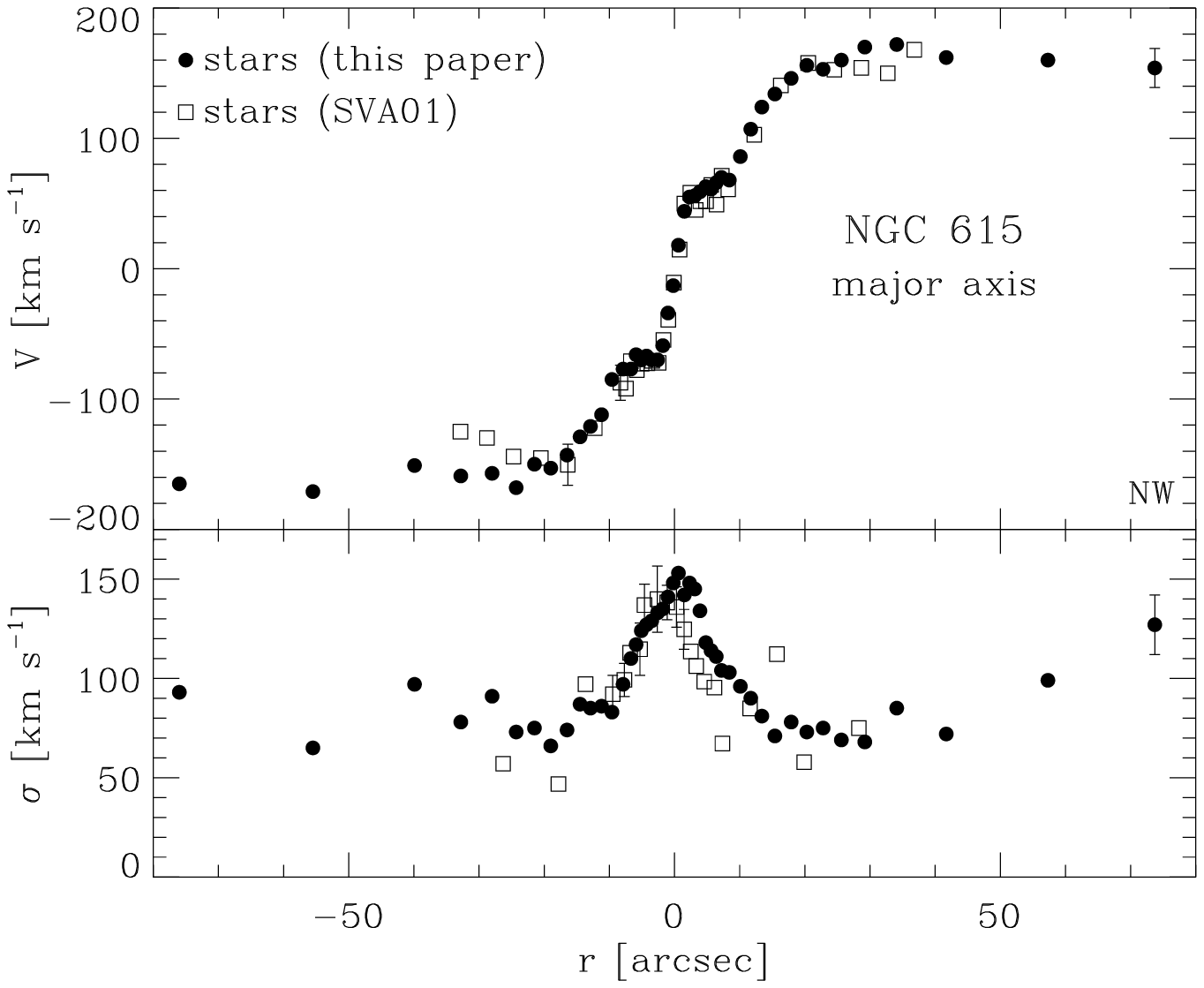}
\end{minipage}
\hspace*{0.5cm}

\begin{minipage}[t]{8.5cm}
\vspace{0pt}
\includegraphics[width=8.5cm]{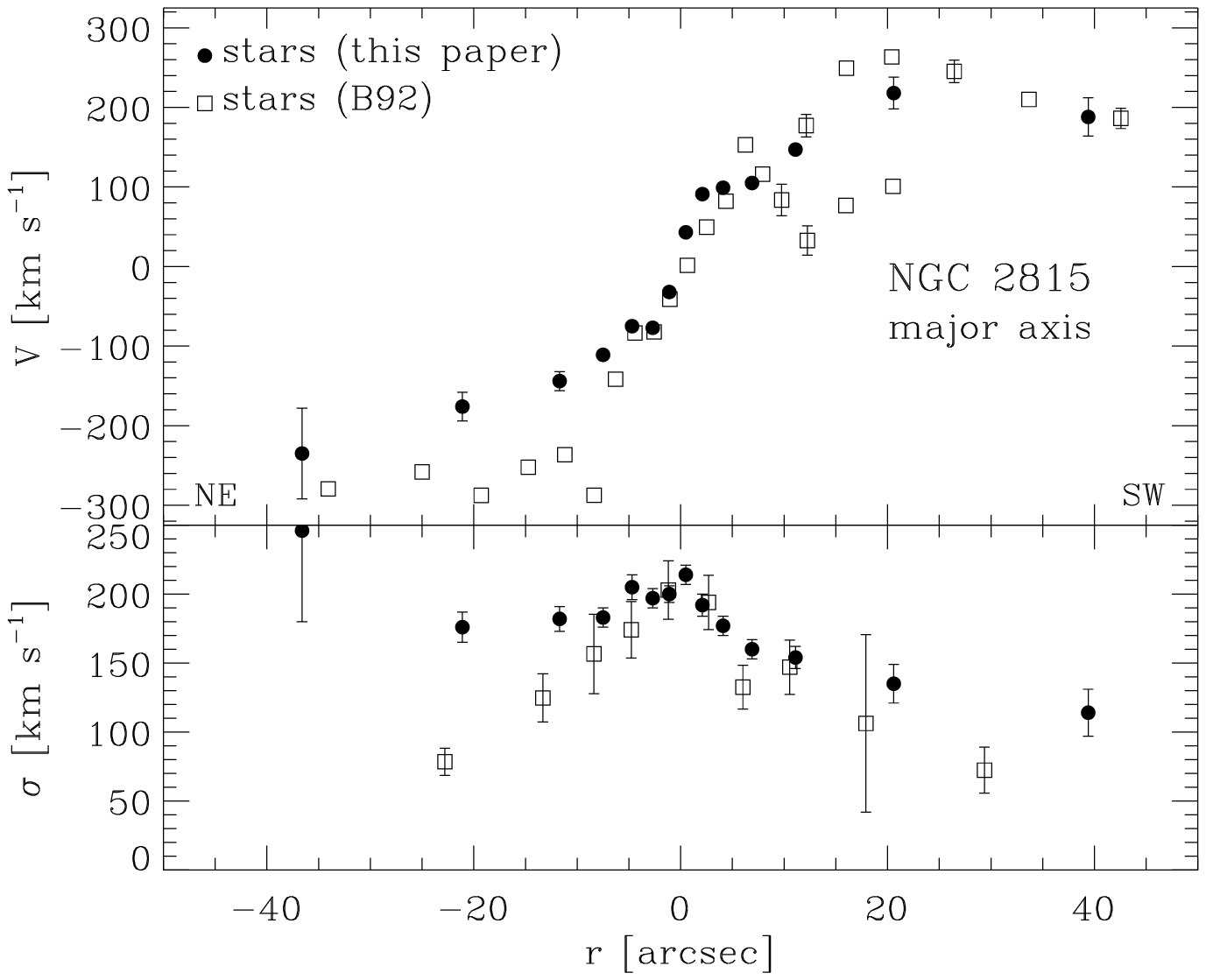}
\end{minipage}
\begin{minipage}[t]{8.5cm}
\vspace{0pt}
\includegraphics[width=8.5cm]{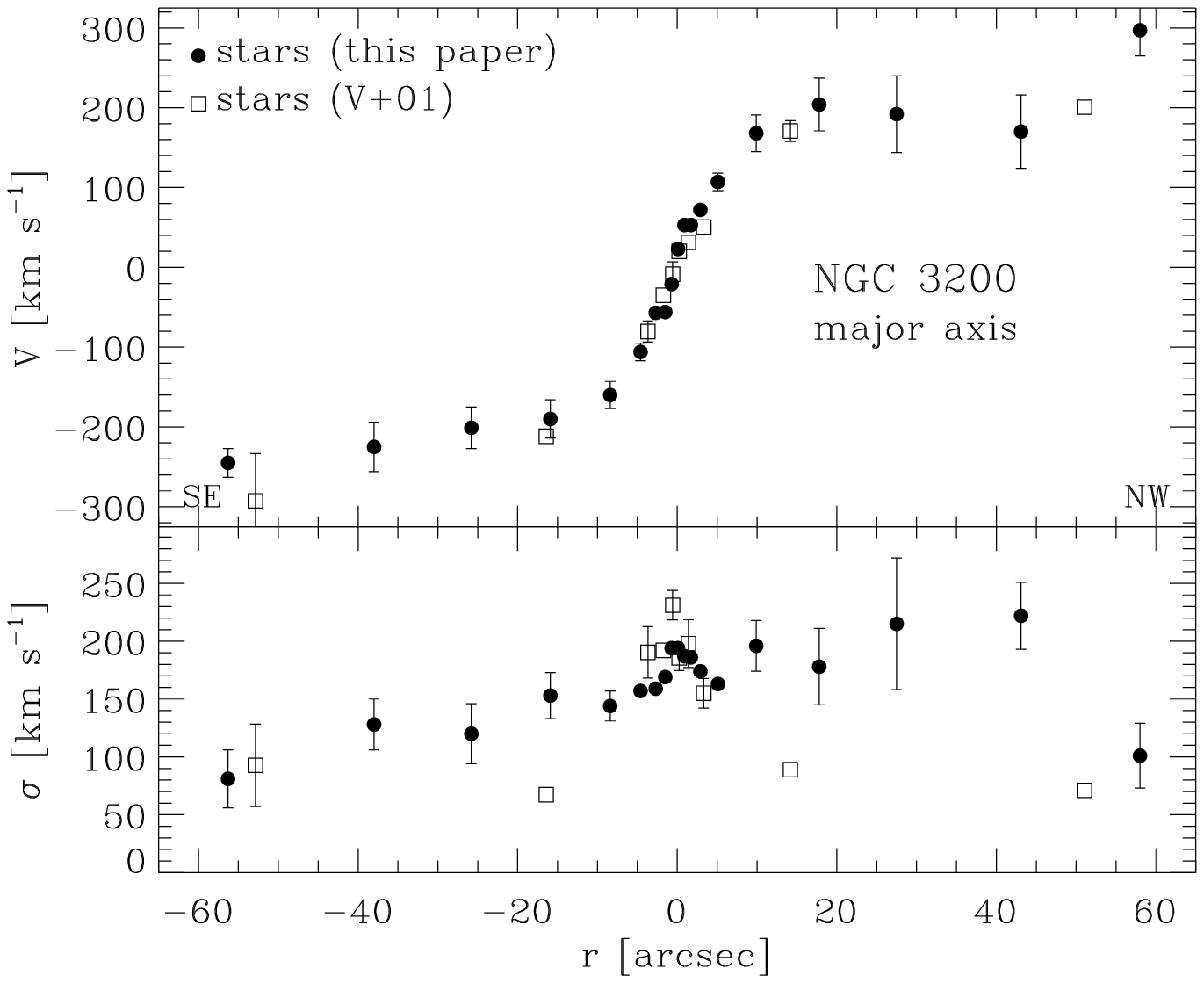}
\end{minipage}
\hspace*{0.5cm}
\caption{The stellar kinematics derived in this study for NGC 615, NGC
  2815 and NGC 3200 compared with those obtained by other authors:
  B92 = Bottema (1992); SVA01 = Sil'chenko, Vlasyuk \& Alvarado
  (2001); V$+$01 = Vega Beltr{\' a}n et al.\ (2001).}
\label{fig:starcomparison}
\end{figure*}

In Fig. \ref{fig:starcomparison} we compared our kinematic
measurements for the stellar component of NGC~615, NGC~2815 and
NGC~3200 to the data derived by Bottema (1992), Sil'chenko, Vlasyuk \&
Alvarado (2001) and Vega Beltr{\' a}n et al.\ (2001), respectively.
The overall agreement is good both for velocities and velocity
dispersions, except for NGC 2815. Indeed, we measured on the south-east
side of this galaxy lower rotation velocities and higher velocity
dispersions with respect to those of Bottema (1992).

\subsection{The central velocity dispersion of stars}
\label{sec:central_sigma}

A tight correlation between the central velocity dispersion
of the spheroid and the galaxy circular velocity has recently been found for a
sample of elliptical and spiral galaxies (Ferrarese 2002; Baes et
al.\ 2003). Since the bulge velocity dispersion and circular velocity
are related to the masses of the central black hole (Ferrarese \&
Merritt 2000; Gebhardt et al.\ 2000) and dark matter halo (e.g.\ Bullock
et al. 2001), respectively, the correlation between velocity dispersion
and circular velocity is equivalent to unity between the mass of the
black hole and the mass of the dark halo (Ferrarese 2002).

With the aim of exploring this correlation in a forthcoming paper
(Pizzella et al.\ 2004, in preparation, but see also Pizzella et
al.\ 2003), we tested how the central stellar velocity dispersion
\ssc\ depends on the radius of the aperture in which it has
been measured. Indeed when comparing the values of \ssc\ to those
derived by other authors it is necessary to take this effect into
account. Ferrarese (2002) took as reference aperture that
corresponding to $R_{\rm e}/8$ where $R_{\rm e}$ is the effective radius of the
bulge, and adopted the aperture correction by Jorgensen, Franx \&
Kjaergaard (1995).

\begin{figure}
   \centering
   \includegraphics[width=12.0cm]{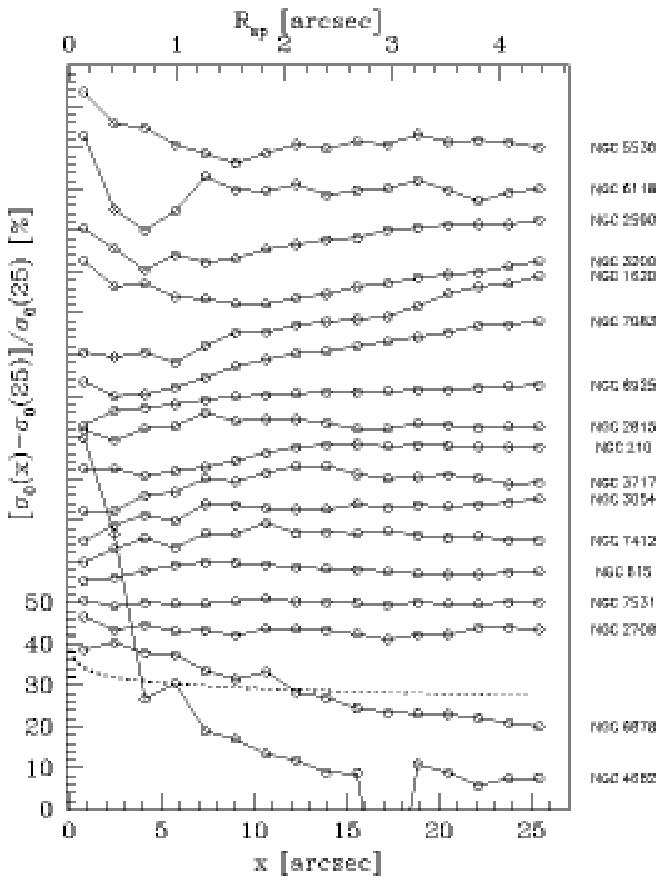}
\caption{Percentage variation in the central velocity dispersion,
  $\sigma_0$, of the sample galaxies as a function of the radial bin,
  $x$ (or of the circularized aperture, $r_{\rm ap}$) in which it has
  been measured. The {\it dotted line\/} is the percentage variation
  obtained for an elliptical galaxy with $\sigma_\ast(0)=150$ \kms\
  and $R_{\rm e}=8''$ by applying the empirical correction by Jorgensen et
  al.\ (1995) .}
\label{fig:test_sigma}
\end{figure}

To test the reliability this correction when applied to spiral
galaxies, we measured  the value of \ssc\ for each sample galaxy in
different radial bins obtained by averaging the central 1, 3, \ldots,
31 rows of the spectrum (corresponding to $x=0\farcs81$, $2\farcs43$,
\ldots, and $25\farcs1$, respectively). In Figure \ref{fig:test_sigma}
the resulting profiles of \ssc\ as a function of the radius of the
circularized aperture are compared to that obtained for a galaxy with
\ssc$=150$ \kms\ and $R_{\rm e}=8\farcs0$ by applying the aperture
correction by Jorgensen et al.\ (1995). These are typical
values for the central velocity dispersions and the bulge effective
radii of our sample galaxies (Baggett, Baggett \& Anderson 1998).

It turns out that in most of cases the observed \ssc\ profile is not
centrally peaked and is nearly constant or even rising for 
large-aperture radii. This suggests that the empirical correction by
Jorgensen et al.\ (1995), which was originally proposed
for elliptical galaxies with centrally peaked velocity dispersions but
negligible rotation velocities, is no longer valid for our late-type
spiral galaxies. In this case the slit aperture is positioned along
the galaxy major axis; therefore, the contribution of the rotation
velocity (which remains spatially unresolved after the binning
process) to the measured velocity dispersion increases with radius.
The contribution of the rotation velocity makes the actual orientation
of the slit is a critical parameter when the aperture correction is
applied.

We therefore decided to not apply any aperture correction to the
measured value of \ssc .  However, as shown in
Figure \ref{fig:test_sigma}, there is no strong dependence of \ssc\
on  aperture for all the sample galaxies, except for NGC 4682.
This peculiar behaviour is due to the low signal-to-noise ratio
($\approx10$) for the innermost two radial bins.

\subsection{Interplay between kinematics of ionized gas and stars}
\label{sec:comparison_gas_star}

Building a complete dynamical model to address the question of the
mass distribution of the sample galaxies is beyond the scope of this
paper.  However, it is possible to derive some hints about their
structure directly from the analysis of the interplay between the
kinematics of their gas and stars. At each radius, $V_\star$ ($\equiv
|v_\star-V_\odot|$) and $V_g$ ($\equiv |v_g-V_\odot|$) are the
observed rotational velocities of the stars and the ionized gas,
respectively. Following Vega Beltr\'an et al. (2001), we distinguished
two classes in the sample galaxies, according to their kinematics and
assuming that the gaseous and stellar discs are coplanar:

\medskip
\noindent
{\it (i)} Galaxies in which ionized gas rotates faster than stars and
has a lower velocity dispersion than the stars (i.e., \Vg$\;>\;$\Vs\
and \sg$\;<\;$\ss): NGC 210, NGC 615, NGC 1620, NGC 2590, NGC 2708
(although it has asymmetric rotation curves for both gas and stars),
NGC 2815, NGC 3200, NGC 3717, NGC 6118, NGC 6878, NGC 6895 and NGC
7083.  The different kinematic behaviour of the gaseous and stellar
components can be explained by a model where the gas is
confined in the disc and supported by rotation, while the stars
are mostly supported by dynamical pressure. This is consistent with the
application of the asymmetric drift (e.g. Binney \& Tremaine 1987).

\medskip
\noindent
{\it (ii)} Galaxies for which \Vg$\;\simeq\;$\Vs\ and
\sg$\;\la\;$\ss: NGC 4682, NGC 5530 and NGC 7412.
The observed difference between the velocity dispersion of the two
components is $\Delta \sigma\leq50$ \kms. The motions of both
ionized gas and stars are dominated by rotation.

The weakly barred spirals NGC 3054 and NGC 7531 show  peculiar
kinematics that are unexpected if the gas is moving in circular
orbits in a disc coplanar to the stellar one. In fact, in NGC 3054
\Vg$\;\leq\;$\Vs\ and \sg$\;<\;$\ss\ for $|r|\la10''$, and
\Vg$\;>\;$\Vs\ and \sg$\;\simeq\;$\ss\ at larger radii, while
NGC 7531 has intermediate properties between the two classes. It is
characterized by \Vg$\;\simeq\;$\Vs\ and \sg$\;<\;$\ss\ for
$|r|\la10''$, and \Vg$\;>\;$\Vs\ and \sg$\;\simeq\;$\ss\ for
$|r|>10''$. These cases cannot be explained without invoking a warp
in the gaseous disc or non-circular gas motions.

\section{The frequency of counterrotation in spiral galaxies}
\label{sec:counterrotation}

\subsection{Detectability of counterrotation}
\label{sec:detectability}

The presence of  counterrotation in disc galaxies is generally
interpreted as the signature of the external origin of the gaseous
component (Bertola, Buson \& Zeilinger 1992). Since we are going to
discuss  the frequency of counterrotation in disc galaxies, it is
necessary to establish our capability of detecting a counterrotating
component when it is present.

In order to estimate the frequency of counterrotating components we
considered the sample of spiral galaxies obtained from Bertola et
al.\ (1996; one Sa galaxy), Corsini et al.\ (1999; six Sa galaxies),
Vega-Beltr\'an et al.\ (2001; sixteen Sa--Scd galaxies. \object{NGC 224} and \object{NGC 3031}
have been excluded from this analysis since the available kinematic
data do not extend to the disc region), Corsini et al.\ (2003; ten
S0/a--Sa galaxies), and this paper (seventeen Sb--Sc galaxies).
The sample consists of 50 spiral galaxies ($0\leq T \leq 6$), with
apparent magnitude $10.1\leq B_T \leq 13.9$, inclination $i \geq 35^\circ $
and systemic velocity $V_\odot \leq 6000$ \kms . Their major-axis
ionized gas and stellar kinematics have been obtained with the same
spatial (FWHM $\approx1''$) and spectral  ($\sigma_{\rm
instr} \approx 50$ \kms ) resolution, and measured with the same analysis
technique. These 50 spiral galaxies represent our qualified sample, for
which we are confident that we are able to detect any counterrotating or
kinematically decoupled gaseous and/or stellar component.

Revealing the presence of ionized gas that is counterrotating with
respect to the stellar component is straightforward. Indeed in most a
cases visual inspection of the shape and orientation of spectral
emission and absorption lines enables us to detect when the gas and stars are
rotating opposite directions (e.g. \object{NGC 4450}; Rubin, Graham
\& Kenney 1992). The techniques we adopted to derive the
gaseous and stellar kinematics allowed us to measure differences of a
few \kms\ in the rotation velocities of the two components. We are therefore
 confident that we are able to detect all the counterrotating gaseous
components hosted in our sample galaxies.

Disentangling a counterrotating stellar component is a more difficult
task. In fact, the kinematics of two counterrotating stellar
populations is measured from the same absorption features and depends
on the dynamical status of the components, as well as on the spectral
resolution of the available data.  We estimated the upper limit on the
fraction of counterrotating stars we can detect assuming that the
counterrotating components are discs with the same scale length and
different luminosity as  follows.
We assumed the stars in the disc of NGC 615 as the prograde component
of the model. We selected the spectrum of NGC 615 since the absorption
lines in the wavelength and radial ranges where we are going to
measure the stellar kinematics are nearly symmetric with respect to
the centre, have a large radial extension and are characterized by a
high signal to noise ($S/N\ga30$ per \AA\ for $|r|>20''$, and
$S/N\geq50$ per \AA\ for $|r|\leq20''$).
We obtained the retrograde component of model by flipping the spectrum of
NGC 615 about the centre of the slit. This corresponds to inverting all
the line-of-sight velocities of the prograde component with respect to
the systemic velocity of the galaxy.
We built a series of synthetic spectra with a different fraction of
counterrotating stars ($f_{\rm CR}= 5\%$, $10\%$,
\ldots, $50\%$) by means of a linear superposition of the prograde and
retrograde models.
The retrograde model has been divided by its own continuum and then
multiplied by the continuum of the prograde model to remove
asymmetries in the continuum shape on the two sides of the galaxy
disc.
We measured the stellar kinematics in the disc-dominated region and
derived the radial profile of LOSVD of the synthetic spectra by
applying the FCQ method as described in Section \ref{sec:measuring}. We
recognized the signature of the presence of two counterrotating
components, namely the splitting of the LOSVD into two peaks, for 
$f_{\rm CR} \geq 15\%$. We assume this value to be our limiting
sensitivity for the detection of counterrotating stars in the galaxy
disc.
We did not consider the bulge-dominated region, where disentangling
the fraction of counterrotating stars requires further constraints on
the dynamical status of the bulge.
We conclude that no more than $15\%$ of the stars in the disc of our
sample galaxies reside in retrograde orbits.

A similar analysis has been done by Kuijken, Fisher \& Merrifield
(1996) who found an upper limit of $5\%$ on counterrotating stars. We
attribute this difference to the fact that they created the prograde and
retrograde models by convolving the spectrum of a kinematic template
with the appropriate LOSVD while we adopted a real (and therefore more
noisy) spectrum.
It has to be noted that this is a limit on the relative surface
brightness of the prograde and retrograde component. If the retrograde
component is more radially concentrated than the prograde one
(e.g. \object{NGC 3593}; Bertola et al.\ 1996), it could be detected even if its
luminosity is less than $15\%$ of the galaxy luminosity.

\subsection{Fraction of disc galaxies hosting a counterrotating gas component}
\label{sec:gas-star}

We found that ionized gas is counterrotating with respect to stars in
2 out 50 galaxies. They are the Sa NGC~3593 (Bertola et al.\ 1996) and
S0/Sa \object{NGC 7377} (Corsini et al.\ 2003), which corresponds to a frequency
of $4\%$. Applying  Poisson distribution statistics, this means
that less than $12\%$ (at the $95\%$ confidence level) of the sample
galaxies have a counterrotating gaseous component.
Only a few other spirals are known to host a counterrotating gaseous
disc (\object{NGC 3626}, Ciri, Bettoni \& Galletta 1996; Haynes et al.\ 2000;
\object{NGC 4138}, Jore, Broeils \& Haynes 1996; Haynes et al.\ 2000; \object{NGC 7217},
Merrifield \& Kuijken 1994) or a kinematically decoupled gas component
(\object{NGC 5854}, Haynes et al. 2000; and the spirals hosting an inner polar
disc listed by Corsini et al.\ 2003).

For comparison, a similar analysis can be done for the lenticular
galaxies. Combining the data of Bertola et al.\ (1992, 1995), Kuijken
et al.\ (1996), and Kannappan \& Fabricant (2001) with those for the S0 \object{NGC 980}
of Vega Beltr\'an et al.\ (2001), we obtained a sample of 53 S0 galaxies
for which the kinematics of the ionized gas and stellar components
have been measured along (at least) the major axis. A counterrotating
or kinematically decoupled gaseous component has been found found in
seventeen of these galaxies. We also took into account  the galaxies whose
ionized gas angular momentum is misaligned with respect to that of the
stellar component since we aimed at deriving the fraction of S0
galaxies hosting a gaseous component of external origin. This fraction
corresponds to $32\%^{+19}_{-11}$ (at the $95\%$ confidence level) from
Poisson distribution statistics and is consistent with previous
results by Bertola et al.\ (1992), Kuijken et al.\ (1996) and Kannappan
\& Fabricant (2001).

\subsection{Fraction of disc galaxies hosting a counterrotating stellar component}
\label{sec:star-star}

In our qualified sample of 50 spiral galaxies we found the presence of
a counterrotating stellar disc in the Sa NGC~3593 (Bertola et
al.\ 1996).
Applying  Poisson distribution statistics this means that less that
$8\%$ (at the $95\%$\ confidence level) of the sample galaxies
have a counterrotating stellar component.
Only two other spirals are known to host two counterrotating stellar
discs, one of these is corotating with the gaseous disc as observed in
NGC 3593. They are the Sa \object{NGC 4138} (Jore et al.\ 1996; Haynes et
al.\ 2000) and the Sab \object{NGC 7217} (Merrifield \& Kuijken 1994). A
kinematically decoupled stellar component orthogonally
rotating with respect to the galaxy disc, has been observed in the
bulge of the early-type spirals \object{NGC 4698} (Corsini et al.\ 1999; Bertola
et al.\ 1999) and \object{NGC 4672} (Sarzi et al.\ 2000).

Kuijken et al.\ (1996) estimated that less than $10\%$ (at the  $95\%$
confidence level) of their S0 galaxies host a significant fraction of
counterrotating stars.
This is actually the case for NGC~4550 (Rubin et al.\ 1992; Rix et
al.\ 1992) where half of the disc stars are moving in retrograde
orbits.

\section{Summary and conclusions}
\label{sec:conclusions}

We have measured the ionized gas and stellar kinematics along the major
axes of seventeen intermediate- to late-type spiral galaxies. The rotation
curves and velocity dispersion profiles of ionized gas and star
typically extend out to $\approx0.9\,R_{25}$ and $\approx0.5\,R_{25}$,
respectively.

In most of cases the different kinematic behaviour of ionized gas and
stars can be easily explained if gas is confined in the disc and
supported by rotation while the stars belong mostly  to the bulge and
are supported by dynamical pressure. However, kinematic peculiarities
have been observed at least in NGC 3054 and NGC 7531.

In addition, we discussed the frequency of counterrotation in disc
galaxies. We considered our seventeen spiral galaxies with those  studied
in previous papers (Bertola et al.\ 1996; Corsini et al.\ 1999, 2003;
Vega Beltr\'an et al.\ 2001) to build a qualified sample of 50 bright
and nearby spirals, ranging from S0/a to Scd, for which the
ionized gas and stellar kinematics have been measured along the major
axis with the same analytical technique.
We found that less than $12\%$ and less than $8\%$ (at the
$95\%$ confidence level) of these galaxies host a counterrotating
gaseous and stellar disc, respectively. For comparison, we found that
$\sim30\%$ of S0s host a counterrotating gaseous disc, and Kuijken
et al.\ (1996) estimated that less than $10\%$ (at $95\%$ confidence
level) host a significant fraction of counterrotating stars.

To interpret the observed frequencies of the gaseous and stellar
counterrotating components of disc galaxies, we suggest a scenario in
which S0 and spiral galaxies are subject to external gas acquisition
with equal probability.

The retrograde acquisition of small amounts of external gas gives rise
to counterrotating gaseous discs only in S0 galaxies, since in spiral
galaxies the acquired gas is swept away by the pre-existing gas.
The formation of counterrotating gaseous discs is favoured in S0
galaxies since they gas-poor systems, while spiral discs host large
amounts of gas (Roberts \& Haynes 1994; Bettoni, Galletta \&
Garc\`{\i}a-Burillo 2003), which is corotating with the stellar
component. When they acquire external gas in retrograde orbits, the
gas clouds of the new retrograde and pre-existing prograde components
collide, lose their centrifugal support, and accrete toward the galaxy
centre. A counterrotating gaseous disc will be observed only if the
mass of the newly supplied gas exceeds that of the pre-existing one
(Lovelace \& Chou 1996; Thakar \& Ryden 1998). A counterrotating
stellar disc is the end-result of star formation in the
counterrotating gas disc.
For this reason we observe a larger fraction of counterrotating
gaseous disks in S0s than in spirals. This also explains why the mass
of counterrotating gas in most S0 galaxies is small compared to that
of the stellar counterrotating components (Kuijken et al.\ 1996).  The
fraction of S0s with a kinematically decoupled gas disc is consistent
with the $50\%$ that we expect if all the gas in S0s is of external
origin (Bertola et al.\ 1992).

Counterrotating gaseous and stellar discs in spirals are both the
results of retrograde acquisition of large amounts of gas, and they are
observed with the same frequency. In this framework stellar
counterrotation is the end result of star formation in the a
counterrotating gaseous disc.

\begin{acknowledgements}
This research has made use of the Lyon-Meudon Extragalactic Database
(LEDA) and of the NASA/IPAC Extragalactic Database (NED). We thank
Sheila Kannappan for providing us her kinematic data and Giuseppe
Galletta for useful discussions.
\end{acknowledgements}

\end{document}